%% file: wp.tex
\documentclass[art11]{article}
\usepackage{graphicx}
\usepackage{amsmath}
\usepackage{caption}
\usepackage{times}
\usepackage{subfigure}
\usepackage{cite}
\usepackage{multirow}
\usepackage{longtable}
\usepackage[affil-sl] {authblk}
\usepackage[ruled,vlined]{algorithm2e}
\usepackage{fancyvrb}

\input{psfig-dvips}

\renewcommand{\baselinestretch}{1.0}

\begin {document}

\title{A Novel Model for Capturing the Multiple Representations during Team Problem Solving based on Verbal Discussions}
\author{{Alex Doboli and Ryan Duke \\
Department of Electrical and Computer Engineering, \\ Stony Brook University, Stony Brook NY 117954-2350 \\ Email: \{alex.doboli, ryan.duke\}@stonybrook.edu}
%\and{Simona Doboli \\
%Department of Computer Science, \\ Hofstra University, Hempstead %NY 11549 \\ Email: \{simona.doboli\}@hofstra.edu
%}
}
\date{}
\maketitle
		
\input{abstract.tex}

%\addtolength{\textwidth}{0.5in} \addtolength{\textheight}{0.8in}
%\addtolength{\topmargin}{-0.2in}
%\addtolength{\oddsidemargin}{-0.2in}
%\addtolength{\evensidemargin}{-0.2in}

%\renewcommand{\baselinestretch}{0.958}
%\documentclass[journal]{}

\thispagestyle{empty}

\input {method}

\bibliographystyle{unsrt}
\bibliography {wp}

\end {document}

%% file: psfig-dvips.tex
\def\PsfigVersion{1.9}
\ifx\undefined\psfig\else \fi

\let\LaTeXAtSign=\@
\let\@=\relax
\edef\psfigRestoreAt{\catcode`\@=\number\catcode`@\relax}
\catcode`\@=11\relax
\newwrite\@unused
\def\ps@typeout#1{{\let\protect\string\immediate\write\@unused{#1}}}
\ps@typeout{psfig/tex \PsfigVersion}

\def\figurepath{./}

\def\@nnil{\@nil}
\def\@empty{}
\def\@psdonoop#1\@@#2#3{}
\def\@psdo#1:=#2\do#3{\edef\@psdotmp{#2}\ifx\@psdotmp\@empty \else
    \expandafter\@psdoloop#2,\@nil,\@nil\@@#1{#3}\fi}
\def\@psdoloop#1,#2,#3\@@#4#5{\def#4{#1}\ifx #4\@nnil \else
       #5\def#4{#2}\ifx #4\@nnil \else#5\@ipsdoloop #3\@@#4{#5}\fi\fi}
\def\@ipsdoloop#1,#2\@@#3#4{\def#3{#1}\ifx #3\@nnil 
       \let\@nextwhile=\@psdonoop \else
      #4\relax\let\@nextwhile=\@ipsdoloop\fi\@nextwhile#2\@@#3{#4}}
\def\@tpsdo#1:=#2\do#3{\xdef\@psdotmp{#2}\ifx\@psdotmp\@empty \else
    \@tpsdoloop#2\@nil\@nil\@@#1{#3}\fi}
\def\@tpsdoloop#1#2\@@#3#4{\def#3{#1}\ifx #3\@nnil 
       \let\@nextwhile=\@psdonoop \else
      #4\relax\let\@nextwhile=\@tpsdoloop\fi\@nextwhile#2\@@#3{#4}}
\ifx\undefined\fbox
\newdimen\fboxrule
\newdimen\fboxsep
\newdimen\ps@tempdima
\newbox\ps@tempboxa
\long\def\fbox#1{\leavevmode\setbox\ps@tempboxa\hbox{#1}\ps@tempdima\fboxrule
    \advance\ps@tempdima \fboxsep \advance\ps@tempdima \dp\ps@tempboxa
   \hbox{\lower \ps@tempdima\hbox
  {\vbox{\hrule height \fboxrule
          \hbox{\vrule width \fboxrule \hskip\fboxsep
          \vbox{\vskip\fboxsep \box\ps@tempboxa\vskip\fboxsep}\hskip 
                 \fboxsep\vrule width \fboxrule}
                 \hrule height \fboxrule}}}}
\fi
\newread\ps@stream
\newif\ifnot@eof       % continue looking for the bounding box?
\newif\if@noisy        % report what you're making?
\newif\if@atend        % %%BoundingBox: has (at end) specification
\newif\if@psfile       % does this look like a PostScript file?
{\catcode`\%=12\global\gdef\epsf@start{%!}}
\def\epsf@PS{PS}
\def\epsf@getbb#1{%
\openin\ps@stream=#1
\ifeof\ps@stream\ps@typeout{Error, File #1 not found}\else
   {\not@eoftrue \chardef\other=12
    \def\do##1{\catcode`##1=\other}\dospecials \catcode`\ =10
    \loop
       \if@psfile
	  \read\ps@stream to \epsf@fileline
       \else{
	  \obeyspaces
          \read\ps@stream to \epsf@tmp\global\let\epsf@fileline\epsf@tmp}
       \fi
       \ifeof\ps@stream\not@eoffalse\else
       \if@psfile\else
       \expandafter\epsf@test\epsf@fileline:. \\%
       \fi
          \expandafter\epsf@aux\epsf@fileline:. \\%
       \fi
   \ifnot@eof\repeat
   }\closein\ps@stream\fi}%
\long\def\epsf@test#1#2#3:#4\\{\def\epsf@testit{#1#2}
			\ifx\epsf@testit\epsf@start\else
\ps@typeout{Warning! File does not start with `\epsf@start'.  It may not be a PostScript file.}
			\fi
			\@psfiletrue} % don't test after 1st line
{\catcode`\%=12\global\let\epsf@percent=%\global\def\epsf@bblit{%BoundingBox}}
\long\def\epsf@aux#1#2:#3\\{\ifx#1\epsf@percent
   \def\epsf@testit{#2}\ifx\epsf@testit\epsf@bblit
	\@atendfalse
        \epsf@atend #3 . \\%
	\if@atend	
	   \if@verbose{
		\ps@typeout{psfig: found `(atend)'; continuing search}
	   }\fi
        \else
        \epsf@grab #3 . . . \\%
        \not@eoffalse
        \global\no@bbfalse
        \fi
   \fi\fi}%
\def\epsf@grab #1 #2 #3 #4 #5\\{%
   \global\def\epsf@llx{#1}\ifx\epsf@llx\empty
      \epsf@grab #2 #3 #4 #5 .\\\else
   \global\def\epsf@lly{#2}%
   \global\def\epsf@urx{#3}\global\def\epsf@ury{#4}\fi}%
\def\epsf@atendlit{(atend)} 
\def\epsf@atend #1 #2 #3\\{%
   \def\epsf@tmp{#1}\ifx\epsf@tmp\empty
      \epsf@atend #2 #3 .\\\else
   \ifx\epsf@tmp\epsf@atendlit\@atendtrue\fi\fi}

\chardef\psletter = 11 % won't conflict with \begin{letter} now...
\chardef\other = 12

\newif \ifdebug %%% turn me on to see TeX hard at work ...
\newif\ifc@mpute %%% don't need to compute some values
\c@mputetrue % but assume that we do

\let\then = \relax
\def\r@dian{pt }
\let\r@dians = \r@dian
\let\dimensionless@nit = \r@dian
\let\dimensionless@nits = \dimensionless@nit
\def\internal@nit{sp }
\let\internal@nits = \internal@nit
\newif\ifstillc@nverging
\def \Mess@ge #1{\ifdebug \then \message {#1} \fi}

{ %%% Things that need abnormal catcodes %%%
	\catcode `\@ = \psletter
	\gdef \nodimen {\expandafter \n@dimen \the \dimen}
	\gdef \term #1 #2 #3%
	       {\edef \t@ {\the #1}%%% freeze parameter 1 (count, by value)
		\edef \t@@ {\expandafter \n@dimen \the #2\r@dian}%
		\t@rm {\t@} {\t@@} {#3}%
	       }
	\gdef \t@rm #1 #2 #3%
	       {{%
		\count 0 = 0
		\dimen 0 = 1 \dimensionless@nit
		\dimen 2 = #2\relax
		\Mess@ge {Calculating term #1 of \nodimen 2}%
		\loop
		\ifnum	\count 0 < #1
		\then	\advance \count 0 by 1
			\Mess@ge {Iteration \the \count 0 \space}%
			\Multiply \dimen 0 by {\dimen 2}%
			\Mess@ge {After multiplication, term = \nodimen 0}%
			\Divide \dimen 0 by {\count 0}%
			\Mess@ge {After division, term = \nodimen 0}%
		\repeat
		\Mess@ge {Final value for term #1 of 
				\nodimen 2 \space is \nodimen 0}%
		\xdef \Term {#3 = \nodimen 0 \r@dians}%
		\aftergroup \Term
	       }}
	\catcode `\p = \other
	\catcode `\t = \other
	\gdef \n@dimen #1pt{#1} %%% throw away the ``pt''
}

\def \Divide #1by #2{\divide #1 by #2} %%% just a synonym

\def \Multiply #1by #2%%% allows division of a dimen by a dimen
       {{%%% should really freeze parameter 2 (dimen, passed by value)
	\count 0 = #1\relax
	\count 2 = #2\relax
	\count 4 = 65536
	\Mess@ge {Before scaling, count 0 = \the \count 0 \space and
			count 2 = \the \count 2}%
	\ifnum	\count 0 > 32767 %%% do our best to avoid overflow
	\then	\divide \count 0 by 4
		\divide \count 4 by 4
	\else	\ifnum	\count 0 < -32767
		\then	\divide \count 0 by 4
			\divide \count 4 by 4
		\else
		\fi
	\fi
	\ifnum	\count 2 > 32767 %%% while retaining reasonable accuracy
	\then	\divide \count 2 by 4
		\divide \count 4 by 4
	\else	\ifnum	\count 2 < -32767
		\then	\divide \count 2 by 4
			\divide \count 4 by 4
		\else
		\fi
	\fi
	\multiply \count 0 by \count 2
	\divide \count 0 by \count 4
	\xdef \product {#1 = \the \count 0 \internal@nits}%
	\aftergroup \product
       }}

\def\r@duce{\ifdim\dimen0 > 90\r@dian \then   % sin(x+90) = sin(180-x)
		\multiply\dimen0 by -1
		\advance\dimen0 by 180\r@dian
		\r@duce
	    \else \ifdim\dimen0 < -90\r@dian \then  % sin(-x) = sin(360+x)
		\advance\dimen0 by 360\r@dian
		\r@duce
		\fi
	    \fi}

\def\Sine#1%
       {{%
	\dimen 0 = #1 \r@dian
	\r@duce
	\ifdim\dimen0 = -90\r@dian \then
	   \dimen4 = -1\r@dian
	   \c@mputefalse
	\fi
	\ifdim\dimen0 = 90\r@dian \then
	   \dimen4 = 1\r@dian
	   \c@mputefalse
	\fi
	\ifdim\dimen0 = 0\r@dian \then
	   \dimen4 = 0\r@dian
	   \c@mputefalse
	\fi
	\ifc@mpute \then
		\divide\dimen0 by 180
		\dimen0=3.141592654\dimen0
		\dimen 2 = 3.1415926535897963\r@dian %%% a well-known constant
		\divide\dimen 2 by 2 %%% we only deal with -pi/2 : pi/2
		\Mess@ge {Sin: calculating Sin of \nodimen 0}%
		\count 0 = 1 %%% see power-series expansion for sine
		\dimen 2 = 1 \r@dian %%% ditto
		\dimen 4 = 0 \r@dian %%% ditto
		\loop
			\ifnum	\dimen 2 = 0 %%% then we've done
			\then	\stillc@nvergingfalse 
			\else	\stillc@nvergingtrue
			\fi
			\ifstillc@nverging %%% then calculate next term
			\then	\term {\count 0} {\dimen 0} {\dimen 2}%
				\advance \count 0 by 2
				\count 2 = \count 0
				\divide \count 2 by 2
				\ifodd	\count 2 %%% signs alternate
				\then	\advance \dimen 4 by \dimen 2
				\else	\advance \dimen 4 by -\dimen 2
				\fi
		\repeat
	\fi		
			\xdef \sine {\nodimen 4}%
       }}

\def\Cosine#1{\ifx\sine\UnDefined\edef\Savesine{\relax}\else
		             \edef\Savesine{\sine}\fi
	{\dimen0=#1\r@dian\advance\dimen0 by 90\r@dian
	 \Sine{\nodimen 0}
	 \xdef\cosine{\sine}
	 \xdef\sine{\Savesine}}}	      
\def\psdraft{
	\def\@psdraft{0}
}
\def\psfull{
	\def\@psdraft{100}
}

\psfull

\newif\if@scalefirst
\def\psscalefirst{\@scalefirsttrue}
\def\psrotatefirst{\@scalefirstfalse}
\psrotatefirst

\newif\if@draftbox
\def\psnodraftbox{
	\@draftboxfalse
}
\def\psdraftbox{
	\@draftboxtrue
}
\@draftboxtrue

\newif\if@prologfile
\newif\if@postlogfile
\def\pssilent{
	\@noisyfalse
}
\def\psnoisy{
	\@noisytrue
}
\psnoisy
\newif\if@bbllx
\newif\if@bblly
\newif\if@bburx
\newif\if@bbury
\newif\if@height
\newif\if@width
\newif\if@rheight
\newif\if@rwidth
\newif\if@angle
\newif\if@clip
\newif\if@verbose
\def\@p@@sclip#1{\@cliptrue}

\newif\if@decmpr

\def\@p@@sfigure#1{\def\@p@sfile{null}\def\@p@sbbfile{null}
	        \openin1=#1.bb
		\ifeof1\closein1
	        	\openin1=\figurepath#1.bb
			\ifeof1\closein1
			        \openin1=#1
				\ifeof1\closein1%
				       \openin1=\figurepath#1
					\ifeof1
					   \ps@typeout{Error, File #1 not found}
						\if@bbllx\if@bblly
				   		\if@bburx\if@bbury
			      				\def\@p@sfile{#1}%
			      				\def\@p@sbbfile{#1}%
							\@decmprfalse
				  	   	\fi\fi\fi\fi
					\else\closein1
				    		\def\@p@sfile{\figurepath#1}%
				    		\def\@p@sbbfile{\figurepath#1}%
						\@decmprfalse
	                       		\fi%
			 	\else\closein1%
					\def\@p@sfile{#1}
					\def\@p@sbbfile{#1}
					\@decmprfalse
			 	\fi
			\else
				\def\@p@sfile{\figurepath#1}
				\def\@p@sbbfile{\figurepath#1.bb}
				\@decmprtrue
			\fi
		\else
			\def\@p@sfile{#1}
			\def\@p@sbbfile{#1.bb}
			\@decmprtrue
		\fi}

\def\@p@@sfile#1{\@p@@sfigure{#1}}

\def\@p@@sbbllx#1{
		\@bbllxtrue
		\dimen100=#1
		\edef\@p@sbbllx{\number\dimen100}
}
\def\@p@@sbblly#1{
		\@bbllytrue
		\dimen100=#1
		\edef\@p@sbblly{\number\dimen100}
}
\def\@p@@sbburx#1{
		\@bburxtrue
		\dimen100=#1
		\edef\@p@sbburx{\number\dimen100}
}
\def\@p@@sbbury#1{
		\@bburytrue
		\dimen100=#1
		\edef\@p@sbbury{\number\dimen100}
}
\def\@p@@sheight#1{
		\@heighttrue
		\dimen100=#1
   		\edef\@p@sheight{\number\dimen100}
}
\def\@p@@swidth#1{
		\@widthtrue
		\dimen100=#1
		\edef\@p@swidth{\number\dimen100}
}
\def\@p@@srheight#1{
		\@rheighttrue
		\dimen100=#1
		\edef\@p@srheight{\number\dimen100}
}
\def\@p@@srwidth#1{
		\@rwidthtrue
		\dimen100=#1
		\edef\@p@srwidth{\number\dimen100}
}
\def\@p@@sangle#1{
		\@angletrue
		\edef\@p@sangle{#1} %\number\dimen100}
}
\def\@p@@ssilent#1{ 
		\@verbosefalse
}
\def\@p@@sprolog#1{\@prologfiletrue\def\@prologfileval{#1}}
\def\@p@@spostlog#1{\@postlogfiletrue\def\@postlogfileval{#1}}
\def\@cs@name#1{\csname #1\endcsname}
\def\@setparms#1=#2,{\@cs@name{@p@@s#1}{#2}}
\def\ps@init@parms{
		\@bbllxfalse \@bbllyfalse
		\@bburxfalse \@bburyfalse
		\@heightfalse \@widthfalse
		\@rheightfalse \@rwidthfalse
		\def\@p@sbbllx{}\def\@p@sbblly{}
		\def\@p@sbburx{}\def\@p@sbbury{}
		\def\@p@sheight{}\def\@p@swidth{}
		\def\@p@srheight{}\def\@p@srwidth{}
		\def\@p@sangle{0}
		\def\@p@sfile{} \def\@p@sbbfile{}
		\def\@p@scost{10}
		\def\@sc{}
		\@prologfilefalse
		\@postlogfilefalse
		\@clipfalse
		\if@noisy
			\@verbosetrue
		\else
			\@verbosefalse
		\fi
}
\def\parse@ps@parms#1{
	 	\@psdo\@psfiga:=#1\do
		   {\expandafter\@setparms\@psfiga,}}
\newif\ifno@bb
\def\bb@missing{
	\if@verbose{
		\ps@typeout{psfig: searching \@p@sbbfile \space  for bounding box}
	}\fi
	\no@bbtrue
	\epsf@getbb{\@p@sbbfile}
        \ifno@bb \else \bb@cull\epsf@llx\epsf@lly\epsf@urx\epsf@ury\fi
}	
\def\bb@cull#1#2#3#4{
	\dimen100=#1 bp\edef\@p@sbbllx{\number\dimen100}
	\dimen100=#2 bp\edef\@p@sbblly{\number\dimen100}
	\dimen100=#3 bp\edef\@p@sbburx{\number\dimen100}
	\dimen100=#4 bp\edef\@p@sbbury{\number\dimen100}
	\no@bbfalse
}
\newdimen\p@intvaluex
\newdimen\p@intvaluey
\def\rotate@#1#2{{\dimen0=#1 sp\dimen1=#2 sp
		  \global\p@intvaluex=\cosine\dimen0
		  \dimen3=\sine\dimen1
		  \global\advance\p@intvaluex by -\dimen3
		  \global\p@intvaluey=\sine\dimen0
		  \dimen3=\cosine\dimen1
		  \global\advance\p@intvaluey by \dimen3
		  }}
\def\compute@bb{
		\no@bbfalse
		\if@bbllx \else \no@bbtrue \fi
		\if@bblly \else \no@bbtrue \fi
		\if@bburx \else \no@bbtrue \fi
		\if@bbury \else \no@bbtrue \fi
		\ifno@bb \bb@missing \fi
		\ifno@bb \ps@typeout{FATAL ERROR: no bb supplied or found}
			\no-bb-error
		\fi
		\count203=\@p@sbburx
		\count204=\@p@sbbury
		\advance\count203 by -\@p@sbbllx
		\advance\count204 by -\@p@sbblly
		\edef\ps@bbw{\number\count203}
		\edef\ps@bbh{\number\count204}
		\if@angle 
			\Sine{\@p@sangle}\Cosine{\@p@sangle}
	        	{\dimen100=\maxdimen\xdef\r@p@sbbllx{\number\dimen100}
					    \xdef\r@p@sbblly{\number\dimen100}
			                    \xdef\r@p@sbburx{-\number\dimen100}
					    \xdef\r@p@sbbury{-\number\dimen100}}
                        \def\minmaxtest{
			   \ifnum\number\p@intvaluex<\r@p@sbbllx
			      \xdef\r@p@sbbllx{\number\p@intvaluex}\fi
			   \ifnum\number\p@intvaluex>\r@p@sbburx
			      \xdef\r@p@sbburx{\number\p@intvaluex}\fi
			   \ifnum\number\p@intvaluey<\r@p@sbblly
			      \xdef\r@p@sbblly{\number\p@intvaluey}\fi
			   \ifnum\number\p@intvaluey>\r@p@sbbury
			      \xdef\r@p@sbbury{\number\p@intvaluey}\fi
			   }
			\rotate@{\@p@sbbllx}{\@p@sbblly}
			\minmaxtest
			\rotate@{\@p@sbbllx}{\@p@sbbury}
			\minmaxtest
			\rotate@{\@p@sbburx}{\@p@sbblly}
			\minmaxtest
			\rotate@{\@p@sbburx}{\@p@sbbury}
			\minmaxtest
			\edef\@p@sbbllx{\r@p@sbbllx}\edef\@p@sbblly{\r@p@sbblly}
			\edef\@p@sbburx{\r@p@sbburx}\edef\@p@sbbury{\r@p@sbbury}
		\fi
		\count203=\@p@sbburx
		\count204=\@p@sbbury
		\advance\count203 by -\@p@sbbllx
		\advance\count204 by -\@p@sbblly
		\edef\@bbw{\number\count203}
		\edef\@bbh{\number\count204}
}
\def\in@hundreds#1#2#3{\count240=#2 \count241=#3
		     \count100=\count240	% 100 is first digit #2/#3
		     \divide\count100 by \count241
		     \count101=\count100
		     \multiply\count101 by \count241
		     \advance\count240 by -\count101
		     \multiply\count240 by 10
		     \count101=\count240	%101 is second digit of #2/#3
		     \divide\count101 by \count241
		     \count102=\count101
		     \multiply\count102 by \count241
		     \advance\count240 by -\count102
		     \multiply\count240 by 10
		     \count102=\count240	% 102 is the third digit
		     \divide\count102 by \count241
		     \count200=#1\count205=0
		     \count201=\count200
			\multiply\count201 by \count100
		 	\advance\count205 by \count201
		     \count201=\count200
			\divide\count201 by 10
			\multiply\count201 by \count101
			\advance\count205 by \count201
		     \count201=\count200
			\divide\count201 by 100
			\multiply\count201 by \count102
			\advance\count205 by \count201
		     \edef\@result{\number\count205}
}
\def\compute@wfromh{
		\in@hundreds{\@p@sheight}{\@bbw}{\@bbh}
		\edef\@p@swidth{\@result}
}
\def\compute@hfromw{
	        \in@hundreds{\@p@swidth}{\@bbh}{\@bbw}
		\edef\@p@sheight{\@result}
}
\def\compute@handw{
		\if@height 
			\if@width
			\else
				\compute@wfromh
			\fi
		\else 
			\if@width
				\compute@hfromw
			\else
				\edef\@p@sheight{\@bbh}
				\edef\@p@swidth{\@bbw}
			\fi
		\fi
}
\def\compute@resv{
		\if@rheight \else \edef\@p@srheight{\@p@sheight} \fi
		\if@rwidth \else \edef\@p@srwidth{\@p@swidth} \fi
}
\def\compute@sizes{
	\compute@bb
	\if@scalefirst\if@angle
	\if@width
	   \in@hundreds{\@p@swidth}{\@bbw}{\ps@bbw}
	   \edef\@p@swidth{\@result}
	\fi
	\if@height
	   \in@hundreds{\@p@sheight}{\@bbh}{\ps@bbh}
	   \edef\@p@sheight{\@result}
	\fi
	\fi\fi
	\compute@handw
	\compute@resv}

\def\psfig#1{\vbox {
	\ps@init@parms
	\parse@ps@parms{#1}
	\compute@sizes
	\ifnum\@p@scost<\@psdraft{
		\special{ps::[begin] 	\@p@swidth \space \@p@sheight \space
				\@p@sbbllx \space \@p@sbblly \space
				\@p@sbburx \space \@p@sbbury \space
				startTexFig \space }
		\if@angle
			\special {ps:: \@p@sangle \space rotate \space} 
		\fi
		\if@clip{
			\if@verbose{
				\ps@typeout{(clip)}
			}\fi
			\special{ps:: doclip \space }
		}\fi
		\if@prologfile
		    \special{ps: plotfile \@prologfileval \space } \fi
		\if@decmpr{
			\if@verbose{
				\ps@typeout{psfig: including \@p@sfile.Z \space }
			}\fi
			\special{ps: plotfile "`zcat \@p@sfile.Z" \space }
		}\else{
			\if@verbose{
				\ps@typeout{psfig: including \@p@sfile \space }
			}\fi
			\special{ps: plotfile \@p@sfile \space }
		}\fi
		\if@postlogfile
		    \special{ps: plotfile \@postlogfileval \space } \fi
		\special{ps::[end] endTexFig \space }
		\vbox to \@p@srheight sp{
			\hbox to \@p@srwidth sp{
				\hss
			}
		\vss
		}
	}\else{
		\if@draftbox{		
			\hbox{\frame{\vbox to \@p@srheight sp{
			\vss
			\hbox to \@p@srwidth sp{ \hss \@p@sfile \hss }
			\vss
			}}}
		}\else{
			\vbox to \@p@srheight sp{
			\vss
			\hbox to \@p@srwidth sp{\hss}
			\vss
			}
		}\fi

	}\fi
}}
\psfigRestoreAt
\let\@=\LaTeXAtSign

%% file: abstract.tex
\begin {abstract}
%\vspace * {-0.1in}
{
Improving the effectiveness of problem solving in teams is an important research topic due to the complexity and cross-disciplinary nature of modern problems. It is unlikely that an individual can successfully tackle alone such problems. Increasing team effectiveness is challenging due to the many entangled cognitive, motivational, social, and emotional aspects specific to teamwork. It is often difficult to reliably identify the characteristics that make a team efficient or those that are main hurdles in teamwork. Moreover, experiments often produced conflicting results, which suggests possibly incorrect modeling of team activities and/or hypothesis formulation errors. Automated data acquisition followed by analytics based on models for teamwork is a intriguing option to alleviate some of the limitations. This paper proposes a model describing an individual's activities during team problem solving. Verbal discussions between team members are used to build models. The model captures the multiple images (representations) created and used by an individual during solving as well as the solving activities utilizing these images. Then, a team model includes the interacting models of the members. Case studies showed that the model can highlight differences between teams depending on the nature of the individual work before teamwork starts. Inefficiencies in teamwork can be also pointed out using the model.       
}
\end {abstract}

%% file: method.tex
\input{intro.tex}

\input{motivation}

\input {related.tex}

\vspace * {-0.1in}
\section {Proposed Modeling Methodology}

\begin{figure}\centering
\includegraphics[width=4.4in]{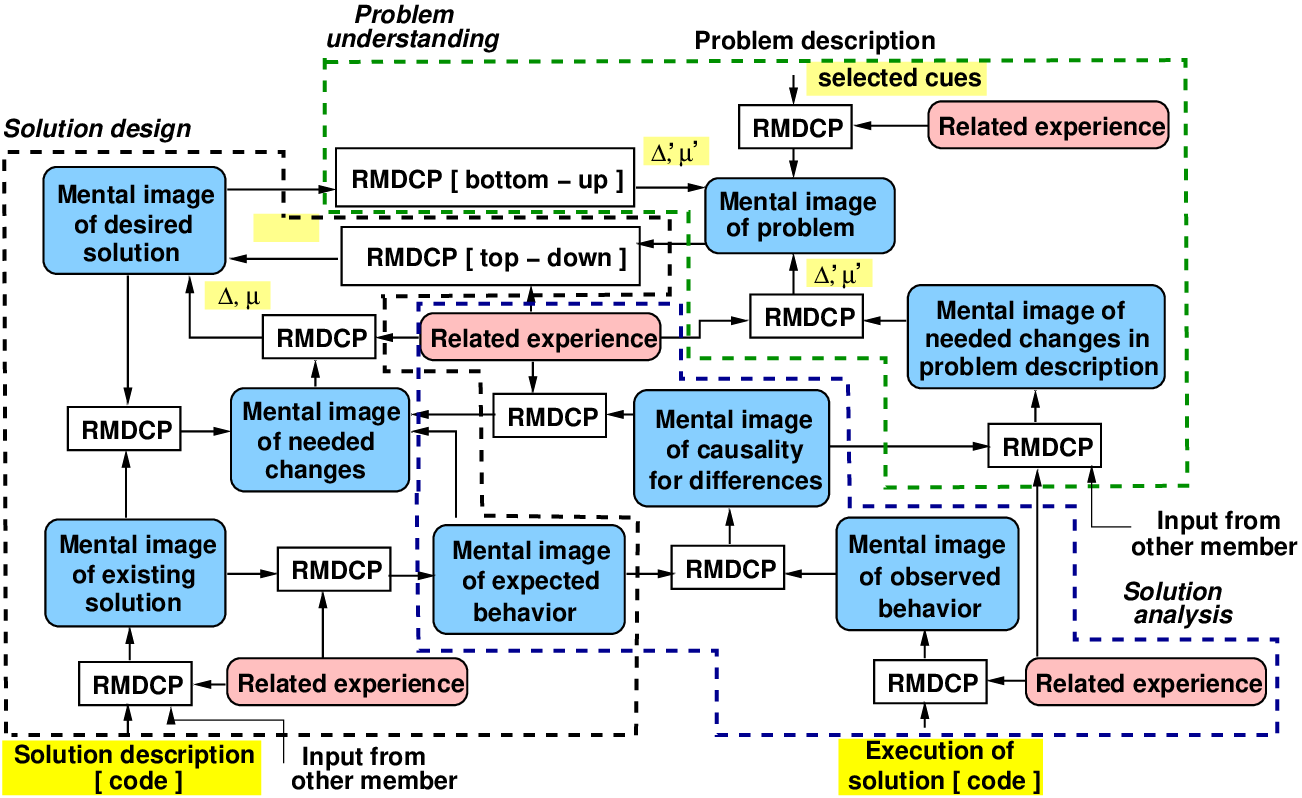}
\caption{Conceptual model for problem understanding and solving}
\label{SAUCE_0}
\vspace * {-0.1in}
\end{figure}
    
This section discusses the proposed modeling methodology to describe problem understanding and solving in teams. The methodology must capture any team activity flow originated for different starting conditions, like having individually completed solutions, individually devised solution fragments, or no available solutions. Besides the overall methodology, the section presents the ten related activities and the overall modeling algorithm. 

\subsection {Modeling Individual Problem Solving}

Problem descriptions include functional (processing) and data requirements, including output values that are expected to result for specific inputs. Unknowns and ambiguities can be part of a problem description. High-level solution descriptions are mixtures of four kinds of description features: (a)~a sequence of actions (e.g., processing), (b)~a set of goals that must be achieved, (c)~a set of requirements for the outputs (i.e. logic conditions that they should meet), and (d)~a set of inputs and corresponding outputs expected for the solutions. It is unclear what parameters influence the mixture of the four feature types. The purpose of the process is to devise a solution that minimizes the difference between the requirements of the problem description and the features of a devised solution.

The model for individual problem solving includes the multiple images used in solving (e.g., the work memory), the nature of knowledge representation (i.e. long-term memory), the individual's goals and expectation about being able to solve the problem and the resulting outcomes, and the sequence of activities over time as part of problem solving. Figure~\ref{SAUCE_0} illustrates the proposed conceptual model for an individual's problem understanding and solving activities. The flow considers two mental images, an image of the problem description and one of the solution. Both images can be mixtures of the above features (a)-(d). The problem image results from the problem understanding activity. The solution image is created by solution analysis. Changes of the images are described by the pair $\Delta', \mu'$, where the first term indicates the change and the second represents the meaning of the change. The changes of the solution images are shown as $\Delta, \mu$. A solution is successfully created when $\Delta, \mu \approx \Delta', \mu' \approx 0$. The difference between the two images is established by RMDCP (Recall - Match - Difference - Combine - Predict) process, which includes four steps: recalling the two images from the memory, matching the images to relate their similar parts, find the difference of the two, and predicting the meaning of the difference in the context of the common matched parts. A bottom-up RMDCP process starting from the image of an existing solution identifies the difference $\Delta',\mu'$ of the problem image, and a top-down RMDCP process starting from the current image of the problem finds the difference $\Delta, \mu$ of the solution image. RMDCP process is executed in the context of the related experience (i.e. existing knowledge) of the member.

A main feature of the model in Figure~\ref{SAUCE_0} is the using of two images connected by the two RMDCPs. Having two images offers the following benefits: (i)~The image of the problem can serve as a reference to set goals for problem solving. It has invariant requirements (even though ambiguities and unspecified elements can be included). Moreover, the image should be abstract enough to support exploration during problem solving. (ii)~The matching and difference computation between the two images serves to guide the decision making during problem solving, such as by $\Delta, \mu \approx \Delta', \mu' \approx 0$, or by maximizing the distance $\Delta, \mu \approx \Delta', \mu'$~\cite{Duke2022b}. (iii)~Matching, difference, and prediction between the two images supports decomposing goals into sub-goals to decide what should be solved next, and also to predict causality, such as how different solution parts contribute to the achieving the problem requirements. Goal decomposition and causality understanding support setting priorities among sub-goals and causal relations, hence guiding decision making during solving process. It also allows to define the difference between the expected and the observed effect of a solution fragment. Finally, the two RMDCPs can express the following two situations: (a)~If they continuously relate the two images then they mimic the subconscious activity that finally creates the feeling that something is right or not, or possibly express sudden insight. (b)~They can represent the conscious activity when a member directs his / her attention on finding the differences between the current solution and problem requirements. 

The description of the resulting differences $\Delta, \mu$ of the top-down RMDCP (which relate the solution to the problem requirements) becomes the current sub-goal of solution design. The solution is then executed. The solution and the execution behavior (like the outputs generated for specific inputs) are inputs to the solution analysis activity. Solution analysis determines differences $\Delta, \mu$ between the expected and observed execution behavior. The differences represent missing or incorrect parts of the solution. The differences modify the mental images of the solution. The analysis can involve reasoning based on the code, such as mentally executing the code. Alternatively, analysis can be based on the data outputs obtained by executing the code for specific data inputs. Reasoning attempts to identify the required change $\Delta, \mu$ to the solution image by following backwards the causal sequence that created the unwanted result. Data-based analysis obtains changes $\Delta, \mu$ by generalizing the input - output behavior  as a rule that creates this behavior, and then combining it with the solution image. Hence, the mental images of a solution, solution design and execution, and analysis form a bottom-up flow that can involve backwards reasoning about the processing steps along the solution's causal relations, and using the differences between the expected and observed inputs and outputs to create the processing rules of the solution.  

\begin{figure}\centering
\includegraphics[width=6.4in]{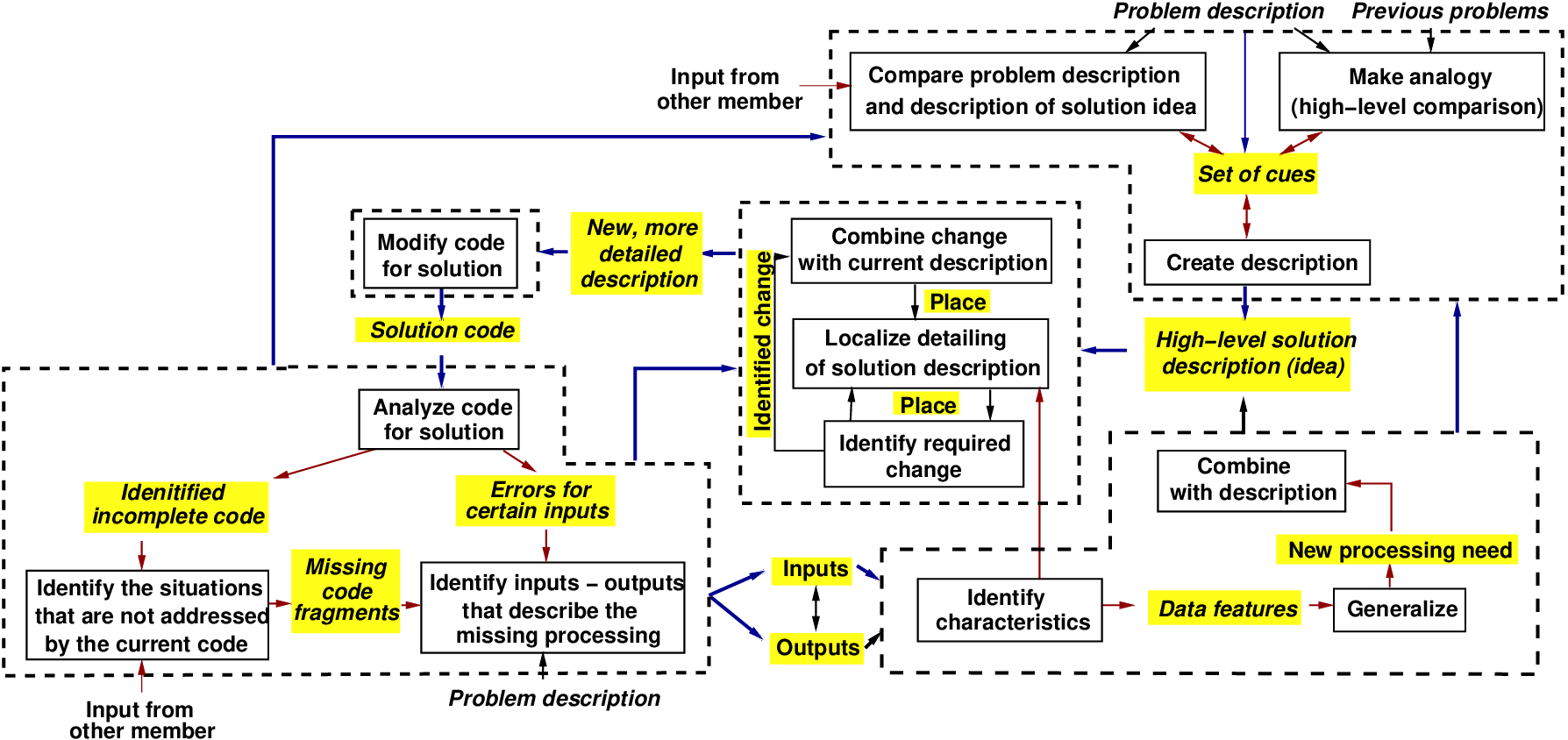}
\caption{Problem understanding and solving dynamics across the semantic space}
\label{SAUCE_2}
\vspace * {-0.1in}
\end{figure}

Figure~\ref{SAUCE_2} depicts the proposed model to describe problem understanding and solving in a team. A high-level solution description of the main idea on how to solve the problem is initially created based on a set of cues selected from the problem description. Alternatively, the process can iterate to refine the solution idea by comparing it with the problem description to expose additional cues that are incorporated into the idea. Another path to create the solution idea is based on analogy with previously solved problems. After relating the new problem description with previously solved, similar problems, an idea for solving the problem is found. This idea is changed to incorporate cues about the differences between the current and previous problems. The relating to previous problems can iterate to identify new cues. 

A new, more detailed description of the solution is devised starting from the high-level solution idea. The iterative detailing process includes three activities, as shown in Figure~\ref{SAUCE_2}, (i)~localizing the place where detailing is performed, (ii)~identifying the change that is performed on the higher-level description, and (iii)~combining the change with the current solution. This process continues until the detailing changes can be transposed into code of the solution.   

After creating or modifying the code, code analysis identifies errors in the code, such as the problem requirements that are not covered or incorrectly realized. It includes identifying the problem requirements that are not covered by the present code and the inputs - outputs for these code fragments. This analysis can continue with the sequence that starts with localizing the corresponding solution detailing (in Figure~\ref{SAUCE_2}) or identifying characteristics for the inputs and outputs that could address the errors, like the expected conditions of the outputs. The identified data features are then combined with the higher-level description. The data features could also lead to changing the analogies used in problem solving. 

\input {team.tex}

\input {case.tex}

\input {conc.tex}

%% file: intro.tex
\section {Introduction}

Problem solving in teams, also called Collaborative Problem Solving (CPS)~\cite{Brennan2010, Fiore2010, Graesser2018}, considers that a team shares a common goal in devising a solution for a problem. Teamwork starts from an initial state of the team, such as team members might have attempted to first individually solve the problem or other situations~\cite{Newell1972, Paulus2013b, Salas2008, Serfaty1998}. Studying CPS has been a major research problem, as modern problems are of a knowledge diversity, complexity, and difficulty that cannot tackled by a single individual~\cite{Graesser2018}. Effective problem solving in teams is not straightforward. Numerous examples show that individuals can often outperform entire teams. 

In a purely mechanistic approach, the problem solving outputs, e.g., ideas, discussions, designs, and social and emotional interactions, would be completely explained by the physiological process in the brains of the team members. Even though this approach might seem extreme, work in neuroaesthetics attempts to explain art along this approach~\cite{Onians2007, Zeki1993}. More recently, some work in engineering design creativity~\cite{Rothmaler2017, Tian2016, Zhao2014} seems to follow the same concept, at least to some degree. While having the potential to offer a causal explanation of high-level cognitive and social activities in terms of physical, physiological signals, it is hard to see how such models scale for complex activities, the well-known curse of dimensionality. The alternative approach, traditionally used in psychology, sociology and organization research, only explores the dependencies among pairs of parameters intuitively selected based on previous experimental observations~\cite{Crosson2010}. These parameters are used to define models and theories, which then spawn new experiments and models. However, while important findings were uncovered using this approach, many experiments report contradicting results and/or suggest conclusions that are subsequently invalidated by new experiments~\cite{Graesser2018}. %There is an obvious gap between the two extremes. 
In our opinion, this limitation relates to insufficient modeling, e.g., model identification, including selecting parameter granularities, model synthesis, like finding parameterized, mathematical expressions of the models, computing the model parameters, and model validation.  

This work argues that a team's ``sauce'', the ingredients that amplify a team's capabilities beyond the sum of its members, is the result of the cognitive, emotional, and social interactions between members, including knowledge access, understanding and extension, and social - emotional changes that these interactions create in the members' behavior during problem solving. In the related literature, joint action represents the interaction among team members that aim to achieve common goals, intentions, or ground~\cite{Brennan2010, Lowe2016, Tomasello2010, Vesper2017}. However, as conflicting experimental results suggest, it is difficult to find what combinations of cognitive, emotional, and social parameters consistently produce effective team problem solving. From a modeling point of view, it can be argued that the reported inconsistencies, like having a positive~\cite{Amabile2005, Bolte2003, Estrada1994, Rowe2007} or a negative impact~\cite{Bjornebekk2008, Dreisbach2004} of positive affect on problem solving, point to modeling errors due to various causes, like incorrect parameter identification or model synthesis. While interactions between members can be directly observed in real-time using smart electronic devices, including those devised by the authors~\cite{Yarden2009, Umbarkar2011, Liu2015, Duke2022a, Duke2022b}, the nature and the evolution of a team model (i.e. its ``sauce'') is hidden. Uncovering the hidden model elements only based on the traditional trial-and-error approach to devise new experiments is cumbersome and unreliable. Instead, there should be more effective methods to uncover the observable and hidden contributions of each member's interactions and behavior to a team, and ultimately to successful team problem solving. The extracted knowledge would enable to create novel ways to form better teams, to analyze team's behavior and performance, and to offer insight for improving problem solving and interactions in a team. Understanding the secret ``sauce'' of teams can also help better team learning in large and diverse settings.

There are important challenges that distinguish team problem solving in real-world settings from other activities, including experiments in laboratory settings. Laboratory experiments usually control one parameter at a time: the outcomes of a reference group (control group) are compared with the outcomes of the experimental group, for which there was an intervention on the studied parameters. Statistical analysis, like ANOVA tests, eliminate any randomness in groups. Then, any observed differences can be assigned to the intervention. The decomposition of complex entanglements between parameters into a sequence of single, independent parameters misses any dependencies between parameters and can be a reason for the observed inconsistencies in experiments. For example, the influence of any previous activities on a current problem decision is hard to capture, even though fixation (when only similar outcomes are offered) and other similar situations are common in real life~\cite{Chrysikou2005, Smith1991, Thomas2009}. Also, problem solving can start from different initial states, like from scratch, reusing previous designs~\cite{Doboli2014, Ferent2013c, Keating2003}, etc., occur in different settings, like in the laboratory, at home, through zoom etc., and span time intervals much longer than the time intervals reported for laboratory experiments. The dependency of cognitive aspects on time, e.g., memory~\cite{Metcalfe1986}, knowledge representations~\cite{Doboli2015, Ferent2013, Ferent2013b, Ferent2013d, Ferent2014, Goel1995, Goldschmidt1997, Hong2004, Jiao2015, Rothmaler2017}, and social and emotional parameters is important~\cite{Bolte2003, Bouty2000, Metcalfe1986, Perry2003}. These elements can arguably make traditional experimental methods inefficient and unreliable. Instead, new experimental approaches are needed, so that they can be seamlessly embedded over a long time into real-world problem-solving situations performed in a variety of conditions.    

This paper proposes a new model describing an individual's activities in a team during problem solving. Models are created based on the verbal discussions between team members during problem solving. The model captures the multiple mental images (representations) created and used by an individual during solving as well as the solving activities that utilize the images. The team model includes the interacting models of the members. The presented model assumes that every team member operates with multiple, individual images of a problem description and related solutions. 
%An image of the solutions fully addresses the image of the 
%problem requirements, if the individual solves the problem. 
Images have the following attributes: (1)~They might not include all details at all time instances, however, any degree of detailing of a problem or solution fragment can be explored during problem solving. (2)~While focusing on specific details, a certain abstract description is created and used for problem solving. (3)~Images might include multiple fragments and ideas representations, which are not necessarily tight together in a unitary description. (4)~Images are connected with an individual's knowledge and previous similar experiences, like previously-solved similar exercises. Under these assumptions, problem solving is the process of filling in the gaps between the images of problem descriptions and solutions using images of solution features (e.g., fragments, characteristics, constraints), input and output features (i.e. examples), and solution goals and sub-goals to be achieved. Features can pertain to different abstraction levels depending on their positioning on the continuum between the images of the problem descriptions and solutions. Hence, in this model, a team's ``sauce'' (defined as the behavior (dynamics) of the cognitive, emotional, and social interactions between members over time) can be seen as the way in which interactions result based on the individual images formed and updated over time during teamwork. For example, there is a matching of the individuals' images, if there is a consensus in a team about a final solution, or there are important differences between the images, if there is disagreement in a team. 

This paper focuses mainly on how cognitive interactions in a team will likely change the individuals' images and produce a certain team behavior. Emotional and social interactions are discussed in other work~\cite{Duke2022a, Duke2022b, Liu2015}. The proposed approach considers that creating a correct solution for which there is an informed consensus of every team member is possible only if each individual's solution image includes a correct understanding of the complete causality (e.g., effects) of each solution feature. Any causal effects missing from an image indicate a possible error in the solution, as that feature was not fully analyzed and agreed on by the team. Then, the considered cognitive interactions include the design (e.g., code), high-level solution descriptions, solution details, cues in explanations, solution places to be changed, identified missing fragments in a solution, suggested changes, errors for certain inputs, inputs and outputs for a certain execution scenario, and data features. Specific cognitive interactions are the result of activities of the general problem-solving flow, such as creating an explanation of a problem or solution, making analogies, analyzing a solution or solution description, combining new ideas with an existing solution description, generalizing from concrete situations, identifying changes to a solution, localizing where changes are needed, modifying a solution, identifying unhandled situations by a solution, finding inputs and outputs for specific errors or missing solution features, and identifying required data processing features. Analyzing the sequence of observed cognitive interactions supports making predictions on the performed solving activities, including finding design decisions executed with more or without fully understanding their consequences at the team level.

This paper presents a novel model and the related methodology to characterize the degree to which cognitive, emotional, and social interactions in a team lead towards correctly solving a problem, while there is an educated consensus among team members about a solution. It is important not only to produce a solution, but also to have the team agree with the solution by understanding and analyzing its features. As different members are likely to have different perspectives, reaching a consensus suggests that the solution withstood the analysis from different points of view. Also, all members reaching a consensus is likely to increase individual learning during problem solving. As an individual's images for problem descriptions and solutions are unknown, predictions are made for their more likely features based on the produced solutions and communicated ideas, explanations, statements, questions, analysis results, agreements, and so on. Each communicated idea serves a certain purpose in solving. It is the result of a response synthesis activity, i.e. detailing, generalization, trial and error, etc., and its communication triggers an understanding process within the specific context and experience of each individual. The methodology uses the type of the observed outputs (code and communications) to estimate the nature and characteristics of the performed activities within a general-purpose problem-solving flow. These activities support then making predictions about the images of the individuals. Predictions about the performed problem-solving activities and individual activities can be used to estimate the quality of the solutions, i.e. by describing the analyzed and missed design opportunities, likely errors as well as design issues that are not supported by consensus or which were not fully understood by all members.

The paper has the following structure. Section~2 offers an experimental motivation of the work. Section~3 discusses related work. Section~4 presents the methodology used to study the factors that describe team interactions, i.e. those defining a team's ``sauce''. Section~4 describes three types of team problem-solving scenarios. Conclusions end the paper.

%% file: motivation.tex
\section {Motivation}

The {\em problem-solving space of a problem description} is the set of solutions and solution fragments that pertain to the semantic space bordered by the problem description and the implementations, e.g., C programs, that correctly address the problem description. Then, problem solving identifies ways of bridging the semantic gap between a problem description and implementations, so that (i)~there is an equivalency between the description and a solution under the assumption that (ii)~any ambiguous, undefined, or other open-ended aspects of the description are consistently identified and addressed. 

{\bf Example}: A problem descriptions requires reading data from a file about date, temperature, humidity, and air pressure. Multiple readings can exist for the same date. Each correct value pertains to a predefined range, but values might be corrupt too. The goal is to find the dates with the most erroneous data readings. In addition, the description might include a set of test data and the associated correct outputs. They add details that clarify uncertainties about the description. Still, the description might remain uncertain with respect to some issues, like if all readings for the same date occur successively in the file, or if they are mixed up with the readings for other dates. 
%It can be argued that many problem descriptions remain ambiguous 
%or undefined despite efforts to clarify all their details. 

\begin{figure}\centering
\includegraphics[width=6.4in]{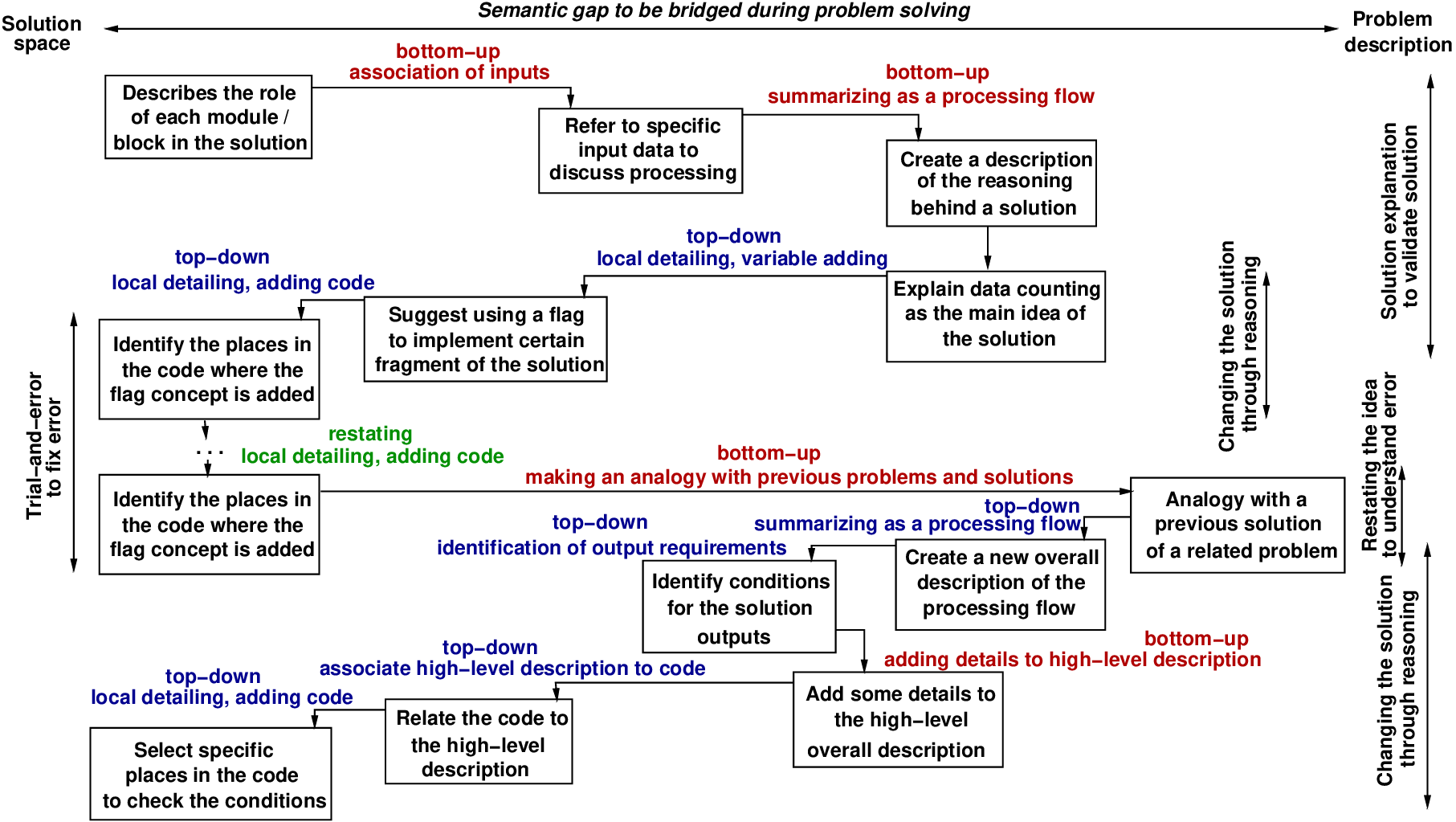}
\caption{Example of problem solving dynamic in a team}
\label{SAUCE_1}
\vspace * {-0.1in}
\end{figure}

The team activity during problem solving can be observed over time to characterize the emerging team dynamic during problem solving~\cite{Doboli2021,Duke2022a,Duke2022b,Liu2015}. As summarized in Figure~\ref{SAUCE_1} for an example, 
%the above problem description was given to a large number of 
%teams of freshmen Electrical and Computer Engineering students 
%as part of their course work requirements for an introductory 
%programming course in C~programming language. 
team discussions reflect the following problem-solving activities and characteristics:
\begin {enumerate}
\vspace * {-0.1in}
\item
Problem solving can be a mixture of activities, which includes not only finding (construction) the actual solution for a description, but also understanding the description, such as reaching some degree of consensus about the problem requirements and constraints, identifying data that helps problem understanding, and solution construction and validation, and reasoning about solutions and their parts. These activities are not separated in distinct, subsequent phases (like it is often assumed in academic work) but are rather mixed-up.   

\vspace * {-0.05in}
\item
Teamwork can have different starting points (initial states) encompassed by the following two extremes: Team members do not understand the problem description to a significant degree, therefore no solutions or solution parts are available and team activity starts by focusing on understanding the problem. The opposite extreme occurs when solutions are already devised to some degree by the individuals, hence teamwork focuses on addressing any of the unsolved issues and solution validation by comparing and harmonizing the solutions by different individuals. As explained later in the paper, the team dynamic is very different in the two cases. Moreover, other starting points can exist, positioned between the two extremes.    

\vspace * {-0.05in}
\item
From a semantic (meaning) point of view, the outputs of an individual during team problem solving pertains to three categories: (i)~Detailing an existing solution fragment, (ii)~abstracting a solution or solution fragment, and (iii) restating an existing fragment. Detailing involves different degrees of added information related to three aspects: (a)~input data used to devise and explain the fragment, (b)~conditions set for the solution results, and (c)~processing steps that form a solution. Abstracting refers to creating descriptions at higher-level of abstractions about the following elements: (a)~conditions of the inputs, (b)~conditions of the output data, and (c)~presentation of the solution procedures and principles. Restating includes producing similar input data, output requirements, and solution steps. Detailing pertains to top-down problem-solving process, and abstraction to a bottom-up process. Detailing, abstraction, and restating belong to the semantic continuum between problem descriptions and solutions. 

\vspace * {-0.05in}
\item
Detailing includes steps on problem description understanding and solution construction with the following features: decomposition of an existing description into new fragments, connection of a new fragment to the rest of the solution, composition of new fragments with each other, adding new variables, capturing data and causal dependencies between variables and fragments, adding new conditions for data and fragments, identification of special and general situations for the solutions and fragments, and identification of new inputs for relevant execution cases. 

\vspace * {-0.05in}
\item
Abstracting refers to problem description understanding and solution development through the following actions: combining local fragments into more overall descriptions, generalizing the outputs obtained for specific data into conditions for solution outputs, and creating higher-level descriptions of a solution or fragment.     
\end{enumerate}      

{\bf Example}: Figure~\ref{SAUCE_1} summarizes the traversal of the problem-solving space as a team attempts to solve the problem. A team of three members was considered. The figure shows the bottom-up activities in red color the, the top-down steps in blue, and the restating in green. The figure also indicates how the individual activities form broader problem-solving phases, like solution explanation to validate its correctness, changing the solution idea through restating, trial-and-error to correct errors in the code, changing (extending) the solution through reasoning, and restating the solution idea to find an error in the code. This example is also discussed in Subsection~4.2. 

In this example, all members attempted to individually solve the problem before meeting in the team setting. However, none of the solutions was complete and correct. In Figure~\ref{SAUCE_1}, the ordering of the performed steps follows the vertical axis from the top to the bottom of the figure. As each of the members already had some ideas on solving the problem, one member started by describing the role (purpose) of each part (i.e. block) in the solution, such as loop statements, conditionals, and so on. The discussion of the detailed processing refers to specific input data and the produced outputs. This step further leads to a description of the high-level reasoning (idea) behind the code. To further clarify the idea, the description is restated by referring to a counting process as being the main concept of the solution. The next step details the broad concept by adding the precise idea of using a flag variable to achieve a certain sub-goal of the broad computation. Solution detailing attempts to identify the specific place where the flag variable should be added. Repeated attempts are made to find the correct insertion place. As the team is unable to correctly add the flag variable to their code, the discussion shifts to making an analogy with a previously solved problem. The analogy helps creating a new overall description of the solution concept, followed by another detailing step that suggests the conditions that the correct outputs must meet. Next, based on the stated output conditions, the high-level solution is modified to include more details. This description is then related to the actual code, such as which modules are expected to complete specific parts of the overall description. This leads to a new attempt to find the C code location that should be modified to extend the existing code.  

As illustrated in Figure~\ref{SAUCE_1}, problem solving is a complex process, which often does not follow a single predefined sequence of steps. It is unclear at this point what specific situations and cues trigger a specific action by a participant, even though there are o few empirical observations, like relying on trial-and-error to find errors or going back to broad, high-level statements about the solution principle after unsuccessful debugging attempts. It can be argued that problem solving relies on a mixture of structured and ad-hoc sequence of decisions in an attempt to find in reasonable time a correct path through the solution space. The gap has an exponential size, and exhaustively exploring all possible code structures, e.g., through automated program generation, could identify a solution but the required time would be prohibitively large.

%% file: related.tex
\section {Related Work}

The state-of-the-art in problem solving in teams (or CPS) has been discussed by a number of classical and recent work in psychology, sociology, and organization research~\cite{Brennan2010, Fiore2010, Newell1972, Graesser2018, Paulus2013b, Salas2008, Serfaty1998}. There is a set of distinguishing characteristics of problem solving in teams as compared to other collaborative activities, like learning, decision-making, and judgement~\cite{Graesser2018}. They include having a shared task, i.e. the problem to be solved, the possibility of individual assessment of the degree to which the solution satisfies the problem (solution quality), role differentiation of team members (including role emergency during problem solving), and the interdependency between members depending on their contribution to successful solving~\cite{Graesser2018}. Problem solving in teams shares a number of advantages with the other collaborative activities, like division of work, multiple sources of information, perspectives and experiences, improved evaluation, and individual simulation by other's ideas, as well as some limitations, such as inefficient communication, social loafing, false information propagation, conflicts, and diffusion of responsibility~\cite{Graesser2018}.

It has been suggested that problem solving in teams comprises of two entangled layers, (i)~the individual layer that includes the emotional and cognitive aspects of each team member, and (ii)~the team layer, which refers to the social elements. The two layers are discussed next.

{\em Individual layer}: This layer refers to an individual's cognition modulated by emotions during problem solving. There are several well-known models about the cognition of problem solving~\cite{Funke2010, Meyer1992}. Among these, our work follows the idea that problem solving finds a path in the problem space from an initial state to the goals of a problem using operators, like evaluation, transformation, and heuristic steps~\cite{Goldschmidt1997, Newell1972}. This model supports the characterization of entire sequences of cognitive activities (e.g., the problem-solving paths), and making prediction on how individual activities are influenced by cognitive, social, and emotional factors. Problem spaces are described as semantic networks, in which nodes represent concepts and links describe actions or relations among concepts~\cite{Holyoak1990}. Then, problem solving is the activation spreading in small steps through the network~\cite{Goldschmidt1997}. The linkage of subsequent design decisions has been a major topic, like identifying the rationale or creating concurrent representations~\cite{Akin1995, Goldschmidt1997}. Network expansion, like using domain space extension~\cite{Rosenman1993}, supports modeling problem solving of ill-defined problems, which require solutions beyond the space of the currently known solutions. Expanding graph-based knowledge representations offers the benefit of expressing features and relations of concepts, which are analogous to properties and connections of building blocks in engineering design~\cite{Jiao2015, Jiao2015c, Jiao2018, Li2018}.

Identifying the problem space representation that reflects the mental (internal) representations is not trivial. Cognitive Task Analysis has been devised in an attempt to characterize the cognitive activities during problem solving based on observations and interviews, process tracing, conceptual techniques, and formal models~\cite{Clark1996, Wei2004}. The problem solving activities are often studied through verbalization~\cite{Conway2002, Kane2004, Decaro2007, Goldschmidt1997}. For example in~\cite{Goldschmidt1997}, the sequence of design steps and their argument are presented for the design of a bicycle rack. The steps used in solving math equations is described in~\cite{Decaro2007}. Even though external solution representations, e.g., sketches, drawings, code, formulas, proofs, verbalization, etc. and other symbolic representations reflect to some degree the internal solution images~\cite{Goldschmidt1997}, there are also important mismatches between the two~\cite{Goel1995, Olson1996}. Related work mainly focuses on mismatches, e.g., physical attributes or negation~\cite{Olson1996}, but there is a much richer set of features that distinguish a concept from its physical realization that operates in a physical setting. This mismatch can have a significant impact on the effectiveness of problem solving, still, cognitive theories have rarely addressed it~\cite{Goel1995}. Experiments also show that multiple solution representation are created during problem solving~\cite{Akin1986, Goldschmidt1997}. For example, figural representation express implementation details while conceptual representations refer to the broader goals and requirements. \cite{Goldschmidt1997} argues that multiple representations convey different kind of information, help memorizing and clarify ambiguities, support interpretation, describe solution transformation, modification and reformulation, and express conflicts and other challenges. 

Studies show that working memory (WM) is a central component in explaining problem solving effectiveness~\cite{Conway2005, Cowan2010, Engle2002}. As explained in~\cite{Decaro2007}, ``WM is a short-term memory system involved in the control, regulation, and active maintenance of a limited amount of information with immediate relevance to the task at hand.'' A similar definition is proposed in~\cite{Conway2005}. WM capacity (WMC) has been shown to be a reliable parameter in discussing other cognitive behaviors, like comprehension and reasoning~\cite{Engle2002}. 
WMC is around 3-5 chunks~\cite{Cowan2001, Cowan2010}. Individual's WMC are assessed through span tasks, like reading, counting, and operation span~\cite{Conway2002, Conway2005, Unsworth2005}. Span tasks require recalling shown letters or words during the problem-solving process, like deciding if a sentence is semantically correct or solving math exercises. Research also focused on the impact of WMC on problem solving under a range affect conditions~\cite{Conway2005, Cowan2010, Unsworth2005}, including reward~\cite{Ashby1999} and work pressure~\cite{Decaro2007}. Experiments show that higher WMC correlates to better problem solving under positive or neutral affect conditions, however, WMC is less important in the case of negative affect situations, as any additional WM capacity is likely used to process negative affects~\cite{Ashby1999, Conway2005}. Dopamine and dopamine paths seem to cause the correlation to positive affect~\cite{Ashby1999,Depue1994}, moods, and emotions~\cite{Derryberry1992}. Similarly, a high WMC can maintain and manipulate more features related to the problem in low pressure situations~\cite{Decaro2007}. However, in high pressure situations, individuals with both high or low WMC relied on shortcuts to solve a problem. WMC is important for story telling the elements (e.g., ideas, concepts) essential to conducting a cognitive activity. For example, as Conway explains for paragraph understanding, creating an integrated image (e.g., stored in a WM chunk) can be realized only if one concurrently holds in mind images about the major premise, the main meaning of the previous sentence, and a fact or opinion discussed in the current sentence~\cite{Cowan2010}. A similar observation was suggested for math problem solving too~\cite{Cowan2001}. It has been argued that problem-solving effectiveness depends on the nature of the information stored in the limited WM space, such as shallow vs. deep features, distractions, or sensory data~\cite{Cowan2010}. Guiding attention to certain features, which are then stored in WMC, was suggested to be significant in problem solving~\cite{Cowan2010}.

Work also shows that affect has an important role in the cognitive activities in problem solving, like goal setting, decision making, and memory recall~\cite{Ashby1999, Dreisbach2004, Isen2001}. For example, positive affect influences the balance between flexibility, e.g., updating WM and switching goals and overcoming fixation~\cite{Chrysikou2005,  Isen2001, Smith1991}, vs. perseverance, i.e. maintaining goals in the presence of distraction, irrelevant answers, and other cues~\cite{Dreisbach2004}. This balance is critical in problem solving, as successful solving involves both considering alternatives and pursuing a certain choice. Affect also changes verbal fluency~\cite{Phillips2002}, helps activating remote associations in memory and producing unusual associations~\cite{Isen1985}, influences the balance between heuristic vs. analytic decision-making~\cite{Isen1983}, categorization~\cite{Isen1984}, and aids implicit judgments of semantic coherence\cite{Bolte2003}. It also influences social categorization~\cite{Isen1992} and job satisfaction~\cite{Estrada1994}. Overviews of computational models for emotions are offered in~\cite{Marsella2010, Ong2020}.

{\em Team layer}: As discussed in Joint Action Theory~\cite{Brennan2010, Lowe2016, Tomasello2010, Vesper2017}, interactions coordinate members through synchronization, entrainment, alignment, and convergence~\cite{Abney2014}, and depend on the conversational context and the team members' intentions and features~\cite{Fusaroli2014}. For example, affiliative conversations have different interaction characteristics than argumentative interactions. Abney et al. explain that coordination is a complex process that occurs at different time-scales and hierarchical levels~\cite{Abney2014}. A complex system behavior emerges because of explicit and implicit matching at phonetic, phonologic, lexical, syntactic, semantic, and situational levels. However, member coordination goes beyond speech attributes. Common ground knowledge, shared visual information, and beliefs about the other team members influence postural and gaze coordination~\cite{Shockley2009}. Affective valuation is also important during interaction~\cite{Lowe2016, Lowe2019}, such as different interpretations of the external stimuli and distinct expectations for the outputs of problem solving. Members must be committed to participate to the team effort. A detailed overview of related models and work is offered in~\cite{Brennan2010}.

Related models in psychology and sociology propose different scenarios for interactions in a team~\cite{Brennan2010}. ``Visual worlds'', arguably the simplest model, assumes that team members follow an active - passive participant approach, with no changing initiative and little coordination between members~\cite{Tanenhaus1995}. The message model states that communication is a probabilistic information flow at a certain rate, during which the sender and receiver must employ the same encoding and decoding of the ``packets of meaning'' represented by words~\cite{Pickering2004}. Social interactions are less important, as there is no coordination, intention recognition, role taking, or matching between the communicated details and the shared common perspective~\cite{Brennan2010}. In contrast, social aspects are addressed in the two-stage model that focuses on lexical entrainment, shared perspective, and reuse of syntactic forms.
Fowler et al. indicate that parsing and speaking durations, speaking rates, turn durations, response latencies, vocal intensities, and accents depend on coordination~\cite{Fowler2007}. The interactive alignment model considers members' perspective adjustment and creation of mental models about other team members~\cite{Horton1996, Keyzar2000}. Lowe et al. propose an extension to Associative Two-Process (ATP) theory to include social cues and affective states~\cite{Lowe2016}. The model uses temporal difference learning to express expectations and to include social cues. Learning considers the magnitude and omission of rewards as well as the temporal difference between stimulus and learned outcome. The work argues not only for the importance of a member's intentional behavior towards a goal but also for the need that an agent learns the other's behavior and then adjusts accordingly. Finally, grounding models emphasize the social, collaborative view, i.e. coordination of meaning, observance of each other, and creation of mental models about others~\cite{Brennan2010, Shockley2009}. 

%Affective (emotional) valuation is also important during team 
%interaction and coordination. 

%% file: team.tex
\subsection {Modeling Team Behavior}              

%The flow in Figure~\ref{SAUCE_2} can cover any specific 
%situation that might occur during problem solving. The overall 
%problem understanding and solving flow in Figure~\ref{SAUCE_2} 
%can generate different sequences of steps (activities). 
There can be a large variety of problem-solving flows, e.g., activity sequences, but flows can be grouped into three main situations depending on the initial status of a team: (a)~when solutions were individually developed before the team activity started, (b)~when there was no available solution before the beginning of teamwork, and (c)~when there were individually-developed solution fragments before the team started to work together, but these fragments did not form a complete solution. The three cases were the only main situations observed in our experiments, even though other situations showed variations of the three cases. The three cases are discussed next followed by details about the team problem-solving model. 

{\em (a)~Solutions were individually developed before team activity starts}: The main goal of the team activity in this case is to integrate the individual solutions into an improved, joint solution for the conjointly accepted problem description, including addressing any misunderstandigs, inefficiencies, and errors that the initial solutions might have. Teamwork includes activities to explain the solutions to the other members, understand someone else's solution, analyze, and compare different solutions to understand their advantages and limitations, combine ideas and fragments of ideas, understand the improvements of the joint solutions compared to the individual solutions, and relate the individual and joint solutions to the problem description. These activities might have to be performed in a context with little incentive to do them, as solutions already exist for the problem. In summary, as shown in Figure~\ref{SAUCE_2}, the flow emphasizes bottom-up activities, like explaining the solutions to others, understanding the solutions of others, comparing solutions to understand their pros and cons, and combining useful solution features. 

{\em (b)~There was no available solution before the beginning of teamwork}: The main goal of the team activity is to jointly clarify any errors, unknowns, and ambiguities by individuals about the problem description and solution ideas, to figure out what these are for each individual, and to stitch together the individual contributions into a correct high-level solution idea and then to correctly detail it until creating the implementation code. Hence, problem solving involves problem decomposition into sub-goals and sub-problems, so that they can be tackled by individuals. It is important to understand the contributions and limits of one's problem understanding and solution, e.g., the uncovered parts of a problem description and the missing elements in a solution idea as compared to the problem description, and how someone else's idea can be combined to advance towards building a complete and correct solution. Coming-up with an overall solution idea based on the individual contributions is also part of the process. Individuals learn new information from others and must integrate it with their own in an attempt to create a complete, higher-level solution idea. Individuals' incentives to solve the problems are high as there is no solution idea available. The problem-solving flow in Figure~\ref{SAUCE_2} emphasizes top-down activities, like decomposing into sub-goals and sub-problems, adjusting them to the knowledge of the participants, and identifying the missing solution fragments that can be solved by individuals.    

{\em (c)~There were individually developed solution fragments before the team starts to work together, but these fragments did not form a complete solution}: The main role is to identify the missing solution parts, and to clarify ambiguities and unknowns in the problem description and to correct errors of the individual, partial solutions. Teamwork is based on explaining solution fragments and understanding the explanations and code. Identifying the missing parts requires localizing problem solving by detailing the existing solutions, identifying the required changes, and combining the changes with the existing solutions. Correcting errors must find their causes through analysis. As high-level descriptions of solutions are likely to exist, problem decomposition into sub-goals and sub-problems is less important than in Case~(b). Incentives to solve the problems are moderately high. The flow in Figure~\ref{SAUCE_2} is a mixture of bottom-up and top-down activities guided by identified missing parts, which become sub-goals for solving. 

Problem solving in a team involves a sequence of cognitive activities performed under individual incentives that are influenced by the team setting. The cognitive activities include an individual's (i)~images used in problem solving (e.g., WM), (ii)~the knowledge representations (i.e. long-term memory), (iii)~the meaning associated to concepts and other outcomes, and (iv)~the decision making that decides the problem-solving steps over time. (v)~Individual expectations modulate the cognitive activities. The team environment includes (vi)~the level of agreement between team members' images, knowledge representations, and their meanings. It also refers to (vii)~a member's flexibility to address the needs of other members as well as (viii)~the expectation about his/her participation to the team's activity. 

A team's effectiveness in the three problem-solving scenarios depends on the following factors specific to each activity:
\begin{enumerate}
\vspace * {-0.15in}
\item
{\em Problem understanding}: The factors describe the degree to which the {\bf requirements} of the problem description and their associated {\bf cues} are correctly and completely considered by a team, and the consistency of the formed mental images of each member (Figure~\ref{SAUCE_0}). The factors include the cues from the problem description that are considered during the development of the high-level solution descriptions (e.g., the completeness of the cue set, cue ordering based on their importance in solving the problem, forgotten cues), the way in which the cues in higher-level descriptions are further tackled during detailing, and the nature of used analogies, including correct and incorrect associations between previous and current problems. The factors also address the degree to which team members agree on the way in which the cues are selected, interpreted, and used to devise the solution ideas across levels of abstraction, including the levels of agreement and disagreement and the importance assigned to cues (e.g., attention). Besides, it presents how considering some cues further influences the inclusion of new cues and cues not in the problem description. Other factors refer to participation, like adaptively raising questions about problem requirements and giving responses, and flexibility in considering alternative interpretations of the requirements. 

\vspace * {-0.05in}
\item
{\em Solution understanding}: Similarly to problem understanding, it refers to the degree to which a solution description is correctly and completely understood by the team members, and the consistency of the related mental images formed by the team members, like the mental images of the {\bf desired} and {\bf existing solutions} (Figure~\ref{SAUCE_0}). Note that the mental images of the desired and existing solutions might only partially {\bf match}, as inconsistencies can occur during encoding in a programming language. Multiple solution descriptions can exist at various abstraction levels. The factors include the degree of understanding of the various solution facets, like the processing flows and their causal influences, the conditions controlling the flow, the handling of special situations, the connections between the higher-level and more detailed descriptions, such as the way in which higher-level parts (including analogies with similar exercises) are realized by lower-level constructs, and the opposite relation between more detailed and more abstract descriptions during abstraction. It also includes the degree to which the individual understanding of the different solution parts is similar or differ. Participation represents a member's engagement in creating a solution, even when the existing solution is not fully understood, as it might have been developed by others. Flexibility describes an individual's willingness to pursue other solution ideas than his/her own. 

\vspace * {-0.05in}
\item 
{\em Problem explanation}: The factors relate to a team member's availability and capability {\bf to restate} the requirements of a problem in response and to satisfy other members' {\bf needs}, like their questions, comments, and doubts. Explanations include alternatives of stating the requirements, logic conditions of the results, and expected outputs for inputs. They can refer to unspecified or ambiguous requirements too. Participation is important, including the availability to adjust explanations depending on expressed or implied needs. Flexibility describes the adjustment of the responses about the problem requirements depending on the specific needs. 

\vspace * {-0.05in}
\item
{\em Solution explanation}: The factors refer to a team member's availability and capability {\bf to describe} the solution ideas at different levels of detailing, including any level between the high-level and code level, depending on the degree to which the explanations {\bf match} the other members' solution understanding {\bf needs}. They include the variety to which different kinds of descriptions are used at various levels of abstraction, like sequences of processing steps, required conditions of the outputs, and relevant inputs and expected outputs. Participation refers to the availability to offer and adjust solution explanations depending on expressed or implied needs. Flexibility describes explanation adjustment based on the needs.  

\vspace * {-0.05in}
\item
{\em Comparison of problem and solution descriptions}: As shown in Figure~\ref{SAUCE_0}, it includes the {\bf RMDCPs} between the mental images of the problem and the desired and expected solutions. The factors include the nature and degree of recall from memory of the features of an image due to the features of the other image, the degree of correct and complete matching of the features of the two mental images, the specifics of the identification of the differences between the two images, the way in which the found differences are integrated with images of previous, similar problems, and the characteristics of the predictions made after integration. The object of participation can be extremely broad, covering all features of a problem description, the high-level and the detailed solutions. Flexibility refers to the possible interpretations of the comparison results, including the importance of similarities and differences.   

\vspace * {-0.05in}
\item
{\em Solution analysis to understand its pros and cons}: Similarly to the previous item, the factors refers to the {\bf RMDCPs} between the mental images of the expected behavior of a solution, the image of the observed behavior, and the image of the causality for the difference between the two (Figure~\ref{SAUCE_0}). The factors include the correctness and completeness of the expected execution traces, output properties, and input - output values, as well as the completeness of the traces, properties, and values of the observed behavior. The correctness, precision, and completeness to which the differences are related to the desired and existing solution images are additional factors too. Participation can be broad, especially to compare the expected and observed behavior. Flexibility is reduced, as the required input - output behavior is fixed by the problem description. The causality of the differences is also fixed by the nature of the solution, even though completely understanding the causality might be hard for complex solutions. Hence, flexibility can include a participant's willingness to accept a presented causality explanation, even if the explanation is not fully understood.    

\vspace * {-0.05in}
\item
{\em Identification of the missing solution fragments}: The features refer to the process of {\bf decomposing} the problem requirements by the {\bf RMDCPs} between the mental images of the problem description, the desired solutions and existing solutions (Figure~\ref{SAUCE_0}). The features describe missing processing, sub-goals and requirements for the missing parts, and input - outputs that should be produced by adding the fragments. They also refer to the completeness, correctness, and the degree of details of the missing parts and their interactions with the parts of an existing solution. 
%The features refer to the completeness, correctness, and the 
%degree of details of the missing parts and their interactions 
%with the parts of an existing solution. 
Participation includes not only sub-goal and missing fragment identification but also analyzing that they reflect well the current solution needs. 

\vspace * {-0.05in}
\item
{\em Identification of the required changes}: The corresponding features relate {\bf reasoning} to find how to address the differences between the images of the desired solutions and existing solutions (Figure~\ref{SAUCE_0}). The RMDCPs predict the outcomes of the differences, and this insight is used to formulate {\bf hypotheses} about the changes that would address the differences, including the necessary detailing of a specific solution part and the abstraction of a part into a higher-level description. The step identifies the places (parts) of an existing solution that must be changed to address the new sub-goals, missing parts, and required modifications (Figure~\ref{SAUCE_2}). Besides, the features describe the degree to which the changes are based on analysis results, like the causality of the differences between the behaviors of the desired and existing solutions, including reasoning, unaddressed situations, and unwanted input - output values. Participation refers to formulating the changes, but also to understanding the correctness of the changes. Flexibility refers to the willingness to accept the change suggestions of others. 
  
\vspace * {-0.05in}
\item
{\em Combining the identified changes with the current solution}: The features refer to {\bf combining} the found changes with the images of desired and existing solutions, and possibly with the image of the problem requirements too, e.g., when ambiguities and unknowns of the requirements are addressed (Figure~\ref{SAUCE_0}). The meaning of the combination is inferred. The features consider the degree to which the behavior of the changes (e.g., changes in processing, and data values and properties) is combined with the behavior of the rest of the solution (Figure~\ref{SAUCE_2}) and the degree to which the combination is consistent across the different mental images (Figure~\ref{SAUCE_0}). Participation to identify all aspects of a combination is important, like the impact on all related variables and processings. Flexibility refers to the degree of acceptance of combination alternatives different than the one proposed by a member.   

\vspace * {-0.05in}
\item
{\em Flow of the problem solving activities}: Features describe the {\bf trace} towards devising a solution, e.g., a working program. The flow of activities in Figures~\ref{SAUCE_0} and~\ref{SAUCE_2} describe how parameters about WM, LTM, participation, and engagement end-up relating the current and previous activities, such as having a sequence of solution detailing activities or switching from solution detailing to analysis. Features present how cues in previous answers, and social and emotional interaction influences the advancement towards the final solution, the breadth and depth of the explored solution space, and the overlapping of the solution ideas of the individuals.     

\end{enumerate}

\begin{figure}\centering
\includegraphics[width=5.0in]{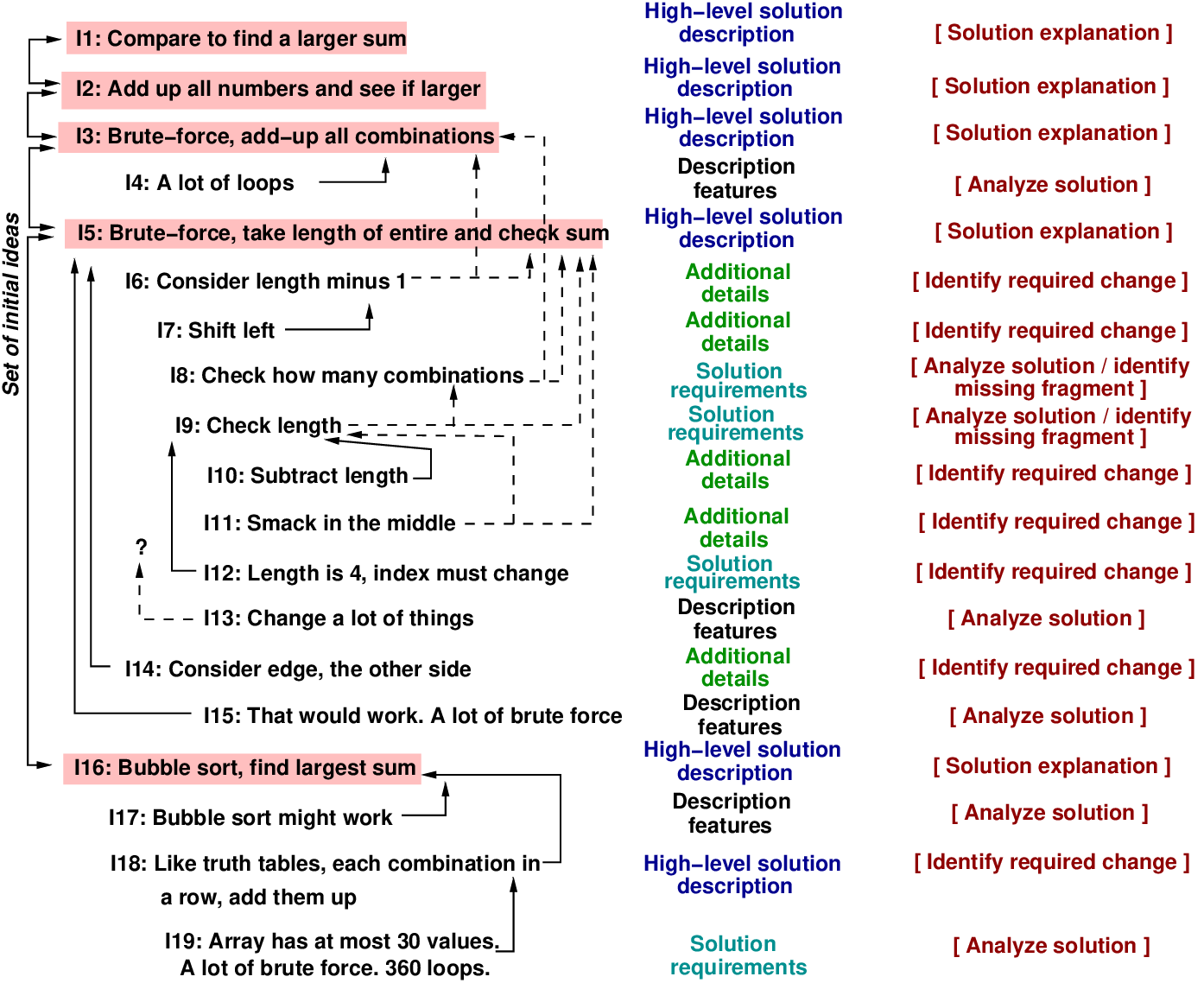}
\caption{Team discussions during problem solving}
\label{SAUCE_3}
\vspace * {-0.1in}
\end{figure}

{\bf Example}: Figure~\ref{SAUCE_3} summarizes a sample of team discussions for a situation in which there were individually developed solution fragments before teamwork commenced, but the fragments did not form a complete solution (Case~(c) in Section~4.2). The tackled problem required the finding of the contiguous sub-array with the largest sum in an existing array with the maximum length of thirty values. The figure shows the inputs produced during problem solving (labeled with letter~{\em I}), their role in problem solving, and the related activity performed by a team member.  

As each team member individually devised fragments of the overall solution, teamwork started with solution explanation by the members (inputs $I_1$ - $I_3$). All were high-level descriptions of an envisioned solution (mental image of the desired solution in Figure~\ref{SAUCE_0}). Note that input $I_1$ was part of the dialog while the team was still discussing the problem description to clarify some doubts (e.g., problem understanding), hence the solving activities were intermingled and not well separated in successive phases. Idea $I_4$ is an analysis of the previous description (idea $I_3$) to communicate a feature of the idea, such as having many loops to implement the idea (i.e. mental image of expected behavior and solution analysis to find pros and cons). Idea $I_5$ is another high-level description (e.g., mental image of desired solution), which, as explained in the next exercise, has a semantic relatedness to the previous inputs $I_1$ - $I_3$. Input~$I_6$ brings additional details by identifying required changes to the description. However, there is a degree of uncertainty to which of the ideas the detailing applies best, e.g., clarifying more the fragment {\em all~combinations} of idea~$I_3$ or referring to {\em length~of~entire} in idea~$I_5$. Also, the specific set words to which it refers is unclear. The uncertainty of the connection was shown with dashed arrows in the figure. Idea $I_7$ restates the meaning of the previous idea $I_6$ by describing one way of implementing the requirement. However, it does not clarify any of the unknowns of idea $I_6$. Input $I_8$ introduces additional ambiguities. Besides the unknown, previous ideas to which it connects, like ideas $I_3$ or $I_5$, its role can be ambiguous too, as it can indicate a concern to be addressed through analysis or identify a missing fragment to be addressed by subsequent solving steps. Similar unknowns exist for idea $I_9$ too. Idea $I_{10}$ indicates a required change to the description to add more details (mental image of needed changes in Figure~\ref{SAUCE_0}). This step clarifies that the previous idea $I_9$ represents a missing fragment of the solution and not an analysis feature, as it could result from interpreting the idea in isolation. Idea $I_{11}$ presents another required change of the description, but it is unclear if it relates to idea $I_5$ or $I_9$. Idea $I_{12}$ adds a new change related to idea $I_9$ (mental images of expected behavior and needed change), thus reinforcing that the latter is a required change too. Idea $I_{13}$ is an analysis of the discussed solution, but it is unknown to which part of the solution it connects to. Idea $I_{14}$ introduces another required change in relation to idea $I_5$. Idea $I_{15}$ is an analysis of the high-level idea $I_5$ as suggested by the word $brute~force$, which designates the meaning of idea~$I_5$. The problem solving in Figure~\ref{SAUCE_3} continues with idea $I_{16}$, which proposes a different solving approach than ideas $I_1-I_3$ and $I_5$. It relies on using an analogy (e.g., bubble sort) with a previous exercise. Input $I_{17}$ is an analysis of the previous idea. Input~$I_{18}$ introduces needed changes to the solution in $I_{16}$. Finally, input $I_{19}$ is an analysis of the detailing introduced by input $I_{18}$. 

Discussions during problem solving mainly focused on aspects related to the mental images of the desired solutions, expected behavior and needed changes, and less about the images of the existing solution, observed behavior, causality of differences between expected and observed behaviors, problem description and needed changes in the problem description. Participation mainly was on adding details to the desired solution and stating some of its cons, but these details were not integrated with each other or with rest of the solution. There was little focus on discussing the missing parts, as there seems to be little attempt to connect the desired solution to the problem description. Without reference to the existing solution and its observed behavior, some ideas (like using bubble sort, $I_{I6}$, or the additions through $I_{11}$, $I_{12}$ and $I_{18}$) are hard to assess about their usefulness. Participation could have been increased by focusing more on the observed behavior and understanding the causality of the differences between the expected and observed behaviors. Regarding flexibility, the team considered two high-level solution alternatives, $I_5$ and $I_{16}$, but it is unclear the degree to which they shaped the individual solution contributions. Encouraging to create complete solutions along different high-level ideas would likely give more insight about the individual flexibility. 

The inputs to the problem solving process pertain to two orthogonal dimensions: the solution space spanned by the high-level descriptions, i.e. the breadth of ideas described by inputs $I_1$, $I_2$, $I_3$, $I_5$, and $I_{16}$, and the solution space described by the details added to the high-level descriptions, such as the specification in depth through identification of missing fragments and required changes. The next example discusses the idea breadth for this team problem-solving example. 

\begin{figure}\centering
\includegraphics[width=6.7in]{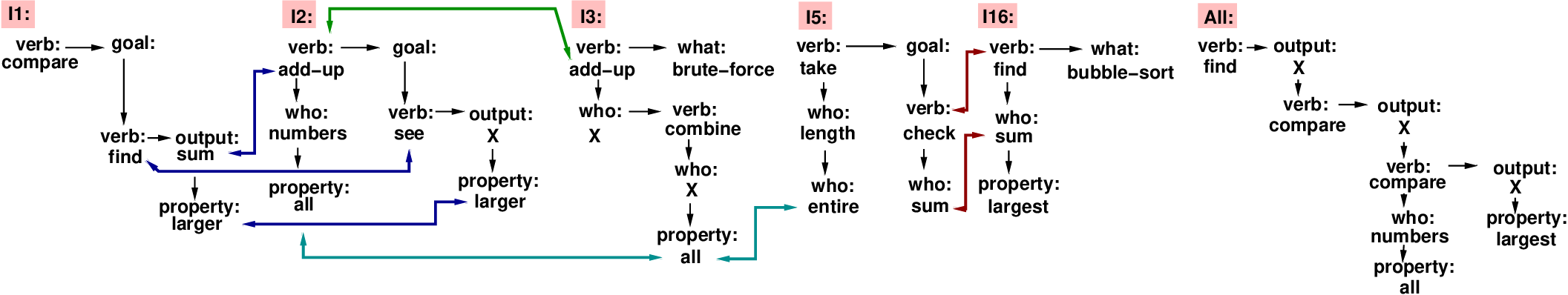}
\caption{Meanings of the descriptions of the alternative high-level solutions}
\label{SAUCE_4}
\vspace * {-0.1in}
\end{figure}

{\bf Example}: Figure~\ref{SAUCE_4} summarizes the meanings of the five descriptions of the high-level solution ideas produced by the team. Each description was expressed by indicating the verb (e.g., actions), the object of the action (i.e. who), the associated goals and outputs, and any related properties of the nouns. In addition, idea~$I_3$ includes a noun (e.g., brute-force) describing an identity (i.e. what). The relatedness of the descriptions can be characterized by the maximum degree of matching between verbs and nouns of similar constructs, like what, where, who, goals, outputs, etc. Some of the components were missing in the inputs, and were marked with letter {\em X}, like the output component of input~$I_2$. Colored arrows show the matched words in different inputs. For example, the blue arrows show the matching of word {\em sum} in input~$I_1$ and word {\em add-up} in input~$I_2$ and that of word {\em larger} in the two inputs. 

The figure illustrates the matchings between consecutive descriptions, even though similar matching can be attempted between any two inputs~$I_i$ and~$I_j$. The matching of descriptions~$I_1$ and~$I_2$ shows that the parts connected through blue arrows express the same meaning in two different structural forms: description~$I_1$ combines two actions (verb {\em find} followed by verb {\em sum}) to create an output, and description~$I_2$ uses a sequence of two actions, in which the second action ({\em see}) is the goal of the first action ({\em add-up}). In addition, verb {\em compare} in idea~$I_1$ and sub-structure {\em all~numbers} in idea~$I_2$ are unmatched. The matching of ideas~$I_2$ and~$I_3$ shows the similarity due to action {\em add-up} in both ideas, as well as the semantic relatedness between fragment {\em all numbers} in idea~$I_2$ and sub-structure {\em all combine (combinations)} in idea~$I_3$. Note that the two are not semantically equivalent, as idea~$I_3$ introduces an additional action through verb {\em combine}, thus it is unmatched in idea~$I_2$. Idea~$I_3$ refers to {\em brute force} to introduce an analogy with similar problems and a way to create the context in which the idea is presented. The matching between ideas~$I_3$ and~$I_5$ is similar to the matching of ideas~$I_1$ and~$I_2$, as ideas~$I_1$ and~$I_5$ have similar structures as do ideas~$I_2$ and~$I_3$. Ideas~$I_1$, $I_2$, and $I_5$ have similar structures that express the goal of an action though a verb. Moreover, the semantics of the sub-structure {\em all~numbers} in idea~$I_2$ is similar to the sub-structure {\em entire~length} in idea~$I_5$ and {\em all} in idea~$I_3$. Finally, the matching of ideas~$I_5$ and~$I_{16}$ includes the words connected through red arrows, while similar to idea~$I_3$, {\em bubble~sort} introduces an analogy with previous solutions. Still, the strict meaning of ideas~$I_5$ and~$I_{16}$ is different, as word {\em largest} is unmatched in idea~$I_5$. Figure~\ref{SAUCE_4} depicts labeled with {\em All} the high-level description that includes all sub-fragments of the five ideas. 

\begin{figure}\centering
\includegraphics[width=5.6in]{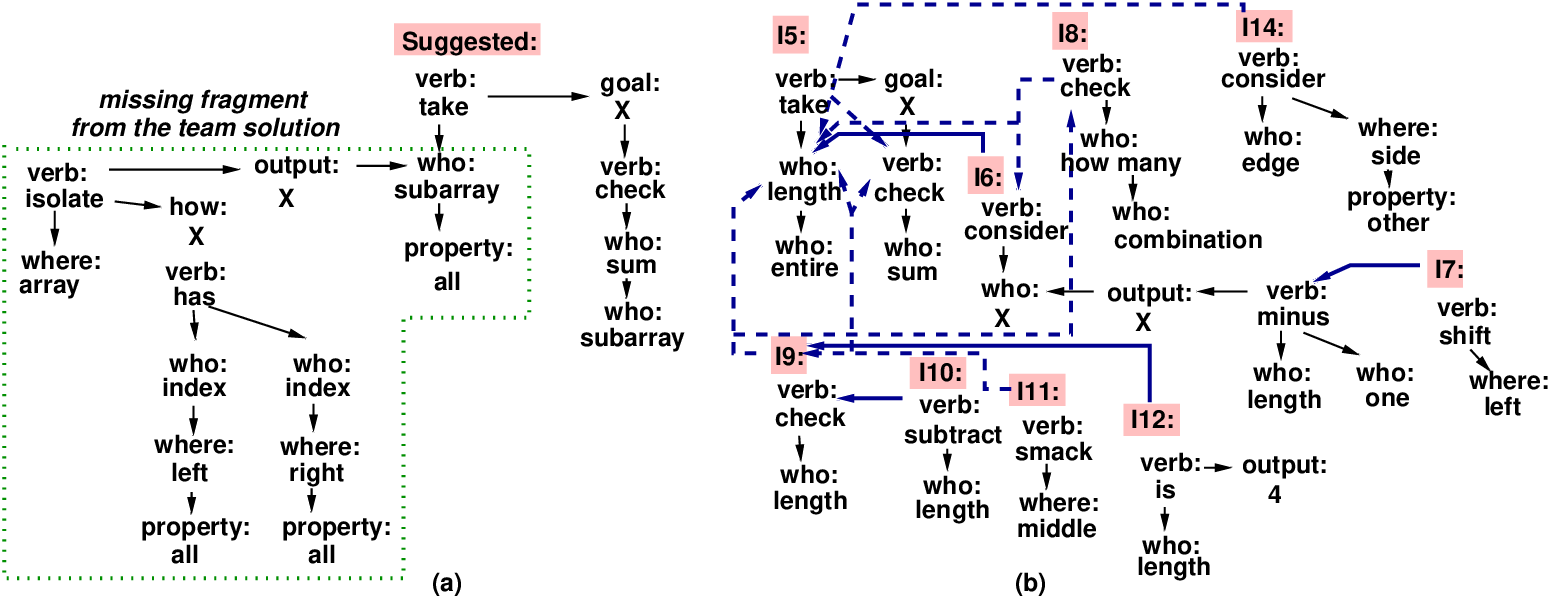}
\caption{(a) Complete solution and (b) team solution after detailing steps}
\label{SAUCE_5}
\vspace * {-0.1in}
\end{figure}
   
As shown in Figure~\ref{SAUCE_0}, detailing high-level ideas used RMDCP steps between mental images to identify differences between desired and existing solution images, identify the details to be added, combine the changes with the current solution, and predict expected outcomes, like pros / cons of the change and comparison to related solutions, including analogies with previously solved problems. The added details are required properties of the variables, requirements of a certain part of the solution, and computations to implement these requirements. They are localized, as they did not change the image of the overall solution idea. It is possible that the meaning of the original and detailed descriptions were not the same, like when new data properties or requirements were added. The added properties and requirements acted as a filter to eliminate options that were not part of a correct solution, an observation that can be used to support the correctness of detailing. From this perspective, a high-level description is like a template (thread) along which detailing is performed, but without changing the template.     

{\bf Example}: Figure~\ref{SAUCE_5}(b) shows the detailing of the high-level idea~$I_5$ in Figure~\ref{SAUCE_4}. Figure~\ref{SAUCE_5}(a) presents a correct solution of the problem. Idea~$I_6$ adds details to the fragment {\em entire~length} of idea~$I_5$. Ideas~$I_5$ and~$I_6$ are semantically not equivalent, as input~$I_6$ describes a changing of variable {\em length}. Even though idea~$I_6$ is not a correct detailing of how variable {\em length} should be handled (as shown in Figure~\ref{SAUCE_5}(a)), it is a step forward in the solving process, as it correctly identifies a missing solution part. Input~$I_7$ is a semantically equivalent restating of idea~$I_6$, but establishing the equivalence is based on observing the equivalence of executing the two operations, {\em minus} one and {\em shift} left (assuming that shifting is over one position of the array). Idea~$I_8$ also refers to fragment {\em entire~length} of input~$I_5$ or to fragment {\em length~minus~1} of idea~$I_6$. Moreover, it can represent an analysis of the solution, or the identification of a missing solution fragment by stating a requirement, like to count the possible combinations of sub-arrays of the overall array. The addition to inspect all sub-arrays is a correct step towards solving the problem (see Figure~\ref{SAUCE_5}(a)), however, the action (i.e. verb {\em count}) is incorrect. Similarly, input~$I_9$ can relate to ideas~$I_5$ or~$I_8$, and describe an analysis or a missing fragment. Idea~$I_{10}$ restates a fragment of idea~$I6$, and defines idea~$I_9$ as having the role to identify a missing solution fragment. Input~$I_{11}$ is ambiguous: its action (e.g., {\em smack}) is undefined with respect to its object (who) and the previous idea it relates to. Input~$I_{12}$ adds a required change related to noun {\em length} of input~$I_9$. Input~$I_{14}$ introduces a required change that relates to input~$I_5$, either its sub-structure {\em entire~length} or its action {\em check}. Figure~\ref{SAUCE_5}(a) indicates the solution fragment that is missing from the solution devised by the team.  

In conclusion, these examples show that the team activity mainly focused on creating and elaborating the mental images of the desired solutions and their expected behavior. Changes were mainly based on matching localized solution fragments, and less on global solution analysis, comparison between the expected and observed behaviors, and matching the observed behavior to the problem requirements. These are possible reasons why the team could not correctly solve the exercise.

%% file: case.tex
\vspace * {-0.1in}
\section {Case Studies}

This section presents case studies for each of the three problem-solving situations depending on a team's status before starting teamwork.
\vspace * {-0.1in}
\subsection {Individually devised solutions existed before starting teamwork}

All team members devised a working solution before meeting in the team setting. As mentioned in Subsection~4.2, the purpose of teamwork should have been to verify the correctness of the solutions with respect to the problem requirements, and to compare the different solution ideas with respect to their pros and cons. An overall observation is that all team discussions were significantly shorter than for the next two problem-solving situations. While all members participated to discussions, the depth of the discussions (e.g., the precision of the expressed ideas) was lower, with fewer and less connected follow-up comments. It is likely that there was less motivation to participate to teamwork as all members already had their own solution, which they assumed to be likely correct. Members usually offered high-level descriptions of their solutions. Descriptions were structured along the processing sequence of a program. Examples based on certain inputs were sometimes used to clarify some steps. However, while the processing steps were individually presented, there was no discussion of how the steps relate to each other, hence often there was no description of the overall solution idea containing the main processing flow of the input data. Subsequent comments by other members were mostly broad and simple, like observing the similarity with previous exercises, discussions about code length and simplicity (which does not require understanding the solution), and certain punctual elements of the solutions. There were few instances in which a question was asked about the code, and the subsequent answers improved the solution. More often, the previous ideas were repeated, which shows little flexibility in understanding the questions. When more detailed descriptions were offered, like comprehensive presentations of the code and variables with reference to certain input examples and correct results, it is likely that keeping track of the offered information was hard for the others.  

{\bf Discussion}: The analysis of team problem solving showed that team members had few incentives to spend time on teamwork, as they already solved individually the problem. With respect to Figure~\ref{SAUCE_0}, there was little interest in performing RMDCPs to form mental images about needed changes, causality of differences, needed changes in problem description, expected behavior, and observed behavior. Teams spent little time to verify the correctness of their solutions with respect to the problem descriptions and to compare the pros and cons of their programs. With respect to Figure~\ref{SAUCE_2}, interaction was less on identifying situations that are not addressed by the current code, identifying inputs - outputs that describe missing processing, comparing problem descriptions and solution ideas, identifying data characteristics, generalizing, and combining with description. There was little flexibility to adjust to the other members' questions.

\subsection {Using individually devised fragments to complete a problem solution}

\begin{figure}\centering
\includegraphics[width=6.4in]{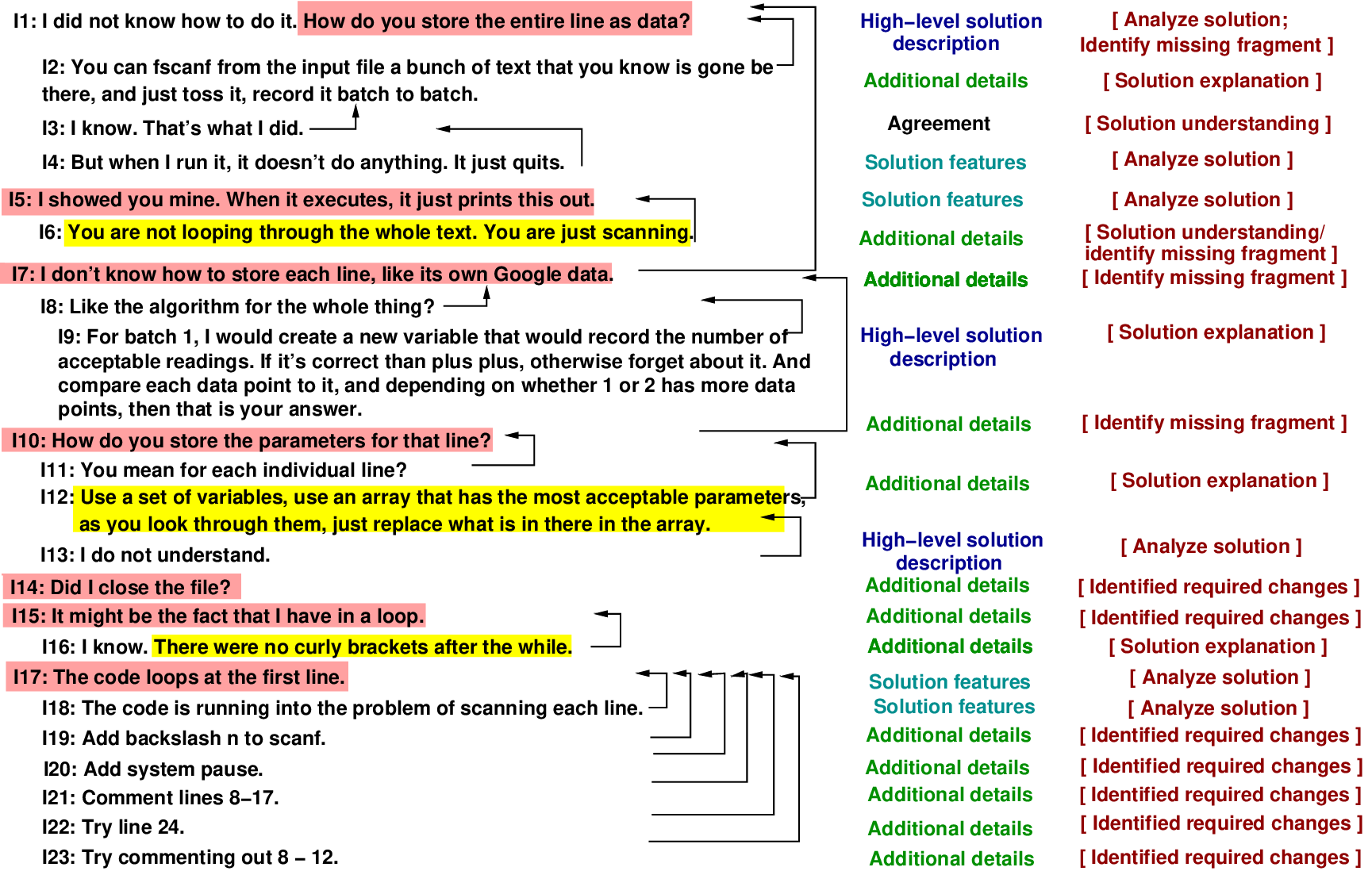}
\caption{Team problem solving dialog starting from individual attempts}
\label{SAUCE_8}
\vspace * {-0.1in}
\end{figure}

Figure~\ref{SAUCE_8} presents discussions during team problem solving for the case in which members attempted to individually solve the problem before meeting in a team setting. Each of the members produced a partial but incomplete solution. In addition to missing fragments, solutions also included errors. The exercise required to ``find and display the batch with the most lines with acceptable parameters, and the batch that has the most lines with unacceptable parameters. Acceptable ranges of parameters for an acceptable seal were as follows: temperature: 150 - 170 C, pressure: 60 - 70 psi, and dwell time: 2 - 2.5 s. A data file named suture.txt contains information about batches. Each line contains the batch number, temperature, pressure and dwell time. Multiple lines can have the same batch number but all lines for the same batch are grouped together''~\cite{Etter2005}. The following discussion refers to the problem solving flow in Figure~\ref{SAUCE_2}.

The team problem-solving part expressed by the dialog in Figure~\ref{SAUCE_8} starts with input~$I_1$ that states a missing (local) fragment from one of the individual solutions. This fragment was found after analyzing the solution and is expressed through a high-level description. The stated missing fragment becomes a sub-goal in problem solving. This sub-goal had to be then addressed in the context of the individual solution. Answer~$I_2$ by another team member does not address the posed sub-goal, but instead explains the entire solution. Details are added to the suggested sequence of steps, like using the C~function {\tt fscanf} together with some more concrete steps, such as separately handling each batch and features of the input data. It is unclear the degree to which acknowledgment~$I_3$ is true considering that response~$I_2$ did not address question~$I_1$. Input~$I_4$ offers a specific feature of the solution, e.g., its execution characteristics found after analyzing the solution. Input~$I_5$ attempts to compare the features of the discussed solution to the features of the solution by another member by explaining that it produces another output. Input~$I_6$ offered by another team member (not the solution author) solves the error mentioned in input~$I_5$. Based on the analysis and understanding of a local part of the code of the solution, the member identified the missing fragment. In addition, the expected purpose of the missing fragment is contrasted to the actual role of the existing code. Hence, input~$I_6$ (highlighted in yellow) is the conclusion of addressing one of the sub-goals, as it fixes one of the errors in the individual solutions.   

Inputs $I_7 - I_{13}$ refocus team problem solving on the initial question~$I_1$ about the missing fragment in one of the solutions. Note that explanation~$I_7$ is incorrect, as the stated details (``like its own Google data'') are not addressing the problem requirements. Still, the other members correctly understood the purpose of the missing fragment. Question~$I_8$ and the subsequent answer~$I_9$ (by the same member that offered input~$I_2$ too) do not answer again the actual question but offer a more detailed explanation of the solution (than explanation~$I_2$). The repeated semantic gap between the question and answers suggests that the discussion is slightly unfocused between the two members. Even though more detailed than description~$I_2$, explanation~$I_9$ is a mixture of description styles, e.g., sequence of broad processing steps and specific details, like using the C~operator {\tt ++} and the referencing to batches one and two. Explanation~$I_9$ includes uncertain parts, like {\tt ``compare each data point to it''}, as the nature of {\tt ``it''} is ambiguous. Still, question~$I_{10}$ suggests that answer~$I_9$ is insufficient, as the question reiterates questions~$I_7$ and~$I_2$ but in a more precise way. Inputs~$I_{11}$ (reinforcement) and then~$I_{12}$ (highlighted in yellow) suggest that the member finally understood the asked question as an explanation was offered, even though it was not understood by the member ($I_{13}$). Answer~$I_{12}$ is unclear as the purpose of using a set of variables or an array with {\tt ``the most acceptable parameters''} is unknown, however, the statement {\tt ``just replace what is in the array''} indicates that the solution might actually refer to separate variables to store the batch numbers with most acceptable and unacceptable values (correct solution). Hence, even though the explanation is incorrect, its underlying idea is arguably correct. 

The next part of team problem solving shifted to other issues than the discussed missing fragment. Inputs~$I_{14}$ and~$I_{15}$ indicate possible changes to the solution. They include details, and act as hypothesis about possible causes that created the observed execution output. Response~$I_{16}$ (highlighted in yellow) addressed the error in the code. It can be argued that the team member narrowed down the search for possible causes until the precise error was found and solved. Therefore, this step describes reasoning about causality, which is important in solution understanding. Inputs~$I_{18} - I_{23}$ attempted to solve the observed code execution features described by input~$I_{17}$. Input~$I_{18}$ offered more details about the features, such as the input characteristics for which the execution feature was observed. Inputs~$I_{19} - I_{23}$ are a sequence of separate hypothesis on how to correct the observed error. They describe causal reasoning by team members to figure out the cause. The randomness of attempts suggests trial-and-error, likely without being guided by an overall image of the solution. 

The analysis of a second team problem-solving example offered similar observations. Team members explained their solutions to each other by using mixed descriptions, like sequences of processing steps presented at a high-level, conditions of the steps, some solution details and special cases, and rarely referring to the code. The descriptions had ambiguous statements. The degree to which the other members correctly understood explanations is unknown. Providing explanations that were understood by others was not trivial. Members often only focused on their solution while paying little attention to understanding other methods. Addressing missing fragments was attempted by starting from specific examples, which while being correctly handled, were hard to generalize for other data too. As they could not identify the reason of an error, the team relied on trial and error to find places where changes were needed. Another incorrect attempt was to modify the code to produce correct results for a particular input, as it did not also produce correct outputs for other inputs. Hence, as backward reasoning was possibly performed for one specific data, the reasoning was not generalized for all data.     

{\bf Discussion}: With respect to the problem-solving flow in Figure~\ref{SAUCE_2}, the main role of teamwork was to identify the missing code fragments and correct the errors of individual solutions. Identifying the missing code fragments involved localizing the solving by detailing an existing solution, identifying the required changes, and combining the changes with the existing solutions. The missing fragments were sub-goals, which were addressed starting from high-level descriptions that had to be connected to the rest of a solution. The process of correcting errors attempted to find their causes, like through backward reasoning. Sequences of hypothesis were set to narrow down to the code part that generated an error. If the process was unsuccessful, the team switched to trial and error. 

Teamwork mainly involved explaining solutions and understanding explanations and code. There was less emphasis on problem description explanation and understanding. Explanations were often mixed descriptions that included sequences of processing activities, features (e.g., conditions) of the input and output data, and examples. Responses contained ambiguous elements and incorrect parts, but it is likely that team members achieved a correct, shared understanding of the intended meaning (despite the actual communication) by correcting it within the context of problem solving to obtain the most useful understanding within that context. There were no questions or comments on the incorrect parts of the explanations. The context set by the individually developed solutions was important for explanation and understanding.     

\subsection {Jointly creating a new solution}

The example presented in Section~4.2 and shown in Figure~\ref{SAUCE_3} illustrates this case. In addition to the discussions on creating a new solution, the team first collaborated on understanding the problem description, as one member had doubts about the meaning of contiguous sub-arrays. The doubt originated in one of the provided test cases, which did not match the member's understanding. The discussion included definitions of terms, and examples to illustrate the using of the definitions and their differences from the provided tests. Errors occurred between the offered examples and the meaning of the problem description. As mentioned in Section~4.2, discussions to clarify the doubts were mixed with high-level ideas of the solutions. Explanations included ambiguous elements. Another requirement was to connect the different explanations offered for the raised doubts. 

A similar problem-solving behavior was observed for another team too. Solving started with making a broad analogy with a previous problem and discussing an input with precise features before offering a high-level description of the processing steps. Solving used examples as a method to identify algorithmic steps to be added to the description. However, there were doubts about the correctness of the additions. Some doubts remained unclarified. Moreover, errors likely occurred when attempting to generalize from the specifics of the examples to code that should process any input. New code was produced based on similarity with previous exercises (analogy), such as a certain code fragment was cut and pasted from an existing solution. It was observed that the team did not understand some of the reused code beyond its expected outputs, hence, mistakes were made due to the differences between the old and new problems. Difficulties also occurred when trying to localize and identify a required change, and then combining it with the existing code. There was little effort for backwards reasoning to understand the cause of an error. Instead, changes were made mainly based on predictions made only by using the inputs and observed outputs (i.e. reactive behavior). Therefore, there is no guarantee (e.g., explanation) that these changes solved the errors besides the fact that the specific expected outputs were obtained. New summaries of the solution were periodically produced, as new code was added. Summaries were high-level sequences of processing steps but were rarely related to the problem description.    

{\bf Discussion}. Team member might have had fragments of the problem solutions, but there were difficulties in forming a complete high-level description. Moreover, there was sometimes a need to discuss a problem description to clarify ambiguities in the descriptions or test data. Devising a complete high-level solution description started with multiple partial ideas of various degrees of semantic similarity. Broad analogies to problems, possibly from other domains, were utilized to create a high-level overall concept. Solution construction required detailing the high-level ideas at various abstraction levels, e.g., the high-level description parts were sometimes mapped directly to code from similar exercises, or successive, more detailed descriptions were created. New detailing was not always semantically equivalent to the previous descriptions, included errors, or were ambiguous with respect to the referred concepts and connected ideas. However, even if erroneous in terms of the expressed processing, they could still have a role in problem solving, like identifying missing fragments. Analysis of alternatives, including their cons, was a team activity. Linking through backward reasoning the observed outputs to their causes was difficult, instead correction was attempted based on the input - output behavior of the code. Finally, problem solving encompassed a part of the solution space, both in breadth and depth. 

After producing a high-level description of the solution, team discussions were localized on adding details to it. Hence, explanation understanding was important to have a broader participation from the entire team. Summarizing code fragments was necessary during team discussions but there was no mechanism beyond using examples to verify the correctness of the summaries or any higher-level descriptions. There was more flexibility between immediate, medium, and long-term contexts. Still, the overall solution image that was formed probably included errors and ambiguities due to uncertain connections between individual ideas.

%% file: conc.tex
\section {Conclusions}

This paper presents a novel model to characterize team activities during devising computer programs that address typical problem requirements. Team members jointly work on solutions and interact through discussions with each other during problem solving. The analysis of problem-solving cases showed that teamwork situations can be grouped into three groups depending on the work executed before meeting in the team set-up: (i)~situations in which team members individually solved a problem before team meetings, (ii)~situations in which team members individually produced incomplete solutions or solutions with significant errors before teamwork started, and (iii)~situations in which no team member could devise a program before meeting in a team. The analysis showed that different kinds of discussions existed in the three situations, suggesting different types of activities being conducted during problem solving. A requirement for the model was to capture the specific characteristics of different problem-solving scenarios observed in real life.      

The proposed team problem-solving model describes that every member operates with multiple mental images of the problem description and solution-related features, like desired and existing solutions, expected and observed behaviors of the solutions, causalities of the differences between expected and observed behaviors, and changes needed to address the differences. The images can be at different levels of abstraction, from high-level ideas to concrete code. They are a mixture of different styles, like sequences of processing steps, expected input - output pairs, or requirements expressed as logic conditions about the results. Descriptions can include unspecified details, ambiguities, unknowns, and errors. The fragments of images are not always tight together in unitary representations. The model connects the various mental images through cognitive activities, like creating problem descriptions, analysis of descriptions, identification of unaddressed issues (e.g., missing functionality and incorrect inputs - outputs), identification of the needed changes, and updating descriptions by combining them with cues and other results of analysis. Analyzing the sequence of team discussions using the proposed model supports making predictions on the performed solving activities, including making design changes without the team fully understanding their consequences. The model supports making predictions about changes of the team members' images because of teamwork and tracking the discussions among them. For example, specific cognitive interactions are the result of certain activities, like explaining a solution to others, analyzing solution correctness, combining new ideas, localizing errors, and so on.

The study of team problem-solving cases using the proposed model showed that discussion characteristics were different for the three situations depending on the nature of the individual solutions before teamwork commencement. Discussions were mostly broad summaries of the solutions if members already devised a program before working in a team. Likely, there was little motivation to have more in-depth descriptions or analysis. If incomplete solutions were devised before teamwork started, then discussions mostly presented alternative solution ideas and detailed certain parts of the ideas. The main effort focused on identifying the missing solution fragments as compared to the problem requirements. The study showed that discussions were not always well structured, hence, the missing parts were not correctly identified. Also, it was unclear the degree to which alternatives were compared to understand their pros and cons, or if reasoning was conducted to determine the missing parts of the solutions. If team members could not design a solution idea before working in a team, then discussions focused on understanding the problem requirements and associating the requirements to previously solved exercises. Explaining ideas was a main part of the discussions, but it was unclear the degree to which explanations were understood by others. Even though members had good a participation in discussions, they rarely followed a systematic procedure to combine individual ideas to progress towards a problem solution. It was unclear how the cues produced by a member illuminated new ideas by others to create joint progress.  

Future work will focus on using the model to devise computational methods to estimate the characteristics of the problem-solving activities and of the mechanisms that connect them in a sequence during teamwork. Another opportunity is to design algorithmic ways of detecting inefficiencies and errors of teamwork, such as ignoring solution analysis and comparison. Finally, the model can be used as a starting point to create new techniques to find correlations between the team members' social and emotional interactions and the properties of the problem-solving flow.